\documentclass{article}
\pdfoutput=1

\usepackage{mathtools}
\usepackage{monash}
\usepackage{multirow,bigdelim}

\usepackage{accents}
\usepackage{placeins}

\definecolor{highlight}{rgb}{0.83, 0.0, 0.0}

\definecolor{decisive}{rgb}{0.83, 0.0, 0.0}
\definecolor{vstrong}{rgb}{0.83, 0.0, 0.0}
\definecolor{strong}{rgb}{0.83, 0.0, 0.0}
\definecolor{substantial}{rgb}{1.0, 0.4, 0.0}
\definecolor{barely}{rgb}{0.18, 0.55, 0.34}

\newcommand{\BR}{\text{BR}}
\newcommand{\LG}{\mathcal{L}}
\newcommand{\pd}{\partial}
\newcommand{\dg}{\dagger}
\newcommand{\modu}[1]{\left\vert{#1}\right\vert}
\mxnewcommand{\ev}{\mathcal{Z}}
\newcommand{\hc}{\text{h.c.}}
\newcommand{\pdarrow}{\accentset{\leftrightarrow}{\pd}^{\mu}}
\newcommand{\pg}[2]{\ensuremath{p(#1\,|\,#2)}}
\mxnewcommand{\like}{\mathcal{L}}
\xnewcommand{\pvalue}{\textit{p}-value}

\mxnewcommand{\invfb}{\,\text{fb}^{-1}}

\newcommand{\rchi}{\text{\raisebox{\depth}{\ensuremath{\chi}}}}

\xnewcommand{\ddcalc}{\texttt{DDCalc-\allowbreak{}1.1.0}\cite{DDCalc:url,Workgroup:2017lvb}}
\xnewcommand{\mo}{\texttt{micrOMEGAs-\allowbreak{}4.3.5}\cite{Belanger:2001fz,Barducci:2016pcb}}

\graphicspath{{figs/}}

\title{\bf Statistical Analyses of Higgs- and $Z$-Portal Dark Matter Models}

\author{{John Ellis$^{1,2,3}$, Andrew Fowlie$^4$, Luca Marzola$^2$} and {Martti Raidal$^{2,3}$} \\
~~\\
$^1$ Theoretical Particle Physics and Cosmology Group, Department of Physics, King's College London, \\
London WC2R 2LS, UK \\
$^2$ National Institute of Chemical Physics and Biophysics, R{\" a}vala 10, 10143 Tallinn, Estonia \\
$^3$ Theoretical Physics Department, CERN, CH-1211 Geneva 23, Switzerland\\
$^4$ ARC Centre of Excellence for Particle Physics at the Tera-scale, Monash University, Melbourne, Victoria 3800, Australia\\
}

\begin{document}

\maketitle

\thispagestyle{empty}

\begin{centering}

{\bf Abstract}

\end{centering}

We perform frequentist and Bayesian statistical analyses of Higgs- and $Z$-portal models of dark matter particles with spin 0, 1/2 and 1. Our analyses incorporate data from direct detection and indirect detection experiments, as well as LHC searches for monojet and monophoton events, and we also analyze the potential impacts of future direct detection experiments.
We 
find acceptable regions of the parameter spaces for Higgs-portal models with real scalar, neutral vector, Majorana or Dirac fermion dark matter particles, and $Z$-portal models with Majorana or Dirac fermion dark matter particles.
In many of these cases, there are interesting prospects for discovering dark matter particles in Higgs or $Z$ decays, as well as dark matter particles weighing $\gtrsim 100$~GeV. Negative results from planned direct detection experiments would still allow acceptable regions for Higgs- and $Z$-portal models with Majorana or Dirac fermion dark matter particles.

\begin{centering}
~~\\
~~\\
~~\\
~~\\
{KCL-PH-TH-58, CERN-TH-2017-246}

\end{centering}

\section{Introduction}

In the last few decades, numerous astrophysical and cosmological observations have confirmed the need for dark matter (DM) via its gravitational interactions, 
with the most precise estimate of its density being provided by Planck satellite measurements of the cosmic microwave
background radiation in combination with other experiments~\cite{Ade:2015xua}. However, many experimental searches 
have failed to find conclusive evidence for dark matter (DM) via non-gravitational interactions. This year alone, 
world-leading upper limits on the strength of DM interactions with matter were set by the PICO~\cite{Amole:2017dex}, XENON1T~\cite{Aprile:2017iyp} and 
PandaX~\cite{Cui:2017nnn} direct detection (DD) experiments, as well as by 
monojet~\cite{Khachatryan:2014rra,Aad:2014nra,Aad:2015zva,Aaboud:2016tnv,Aaboud:2017dor} and monophoton~\cite{Aad:2014tda,Aaboud:2016uro} 
searches for DM at the Large Hadron Collider (LHC). Thus, to date the experimental evidence for DM remains limited to its gravitational interactions.
 
In many models DM is composed of weakly-interacting massive particles (WIMPs) that were in thermal equilibrium with Standard Model (SM)
particles in the early Universe. In such a case, the DM relic abundance is predicted to be that at which WIMPs freeze out from thermal 
equilibrium in the early Universe. 
WIMP decays are typically forbidden by a symmetry, \eg by $\mathbb{Z}_{2,3}$~\cite{Ibanez:1991pr,McDonald:1993ex,Belanger:2012zr} or more generally by $\mathbb{Z}_N$~\cite{Dreiner:2005rd,Belanger:2014bga}, 
and the predicted WIMP abundance agrees `miraculously' with the astrophysical and cosmological observations if their annihilation cross section is that of
the weak interactions in the SM. The simplest WIMP models add a single particle to the SM --- the WIMP itself --- whereas more involved frameworks propose a whole new sector of particles, the lightest of which provides the DM. In the majority of the embedding models, 
the WIMP is a scalar, a fermion or a vector particle, whose annihilations into SM particles may occur through a Higgs boson or $Z$-boson portal.

Assuming that the DM is an SM singlet, gauge-invariant renormalizable portals are possible for scalar or vector DM interacting via a Higgs portal. 
In the remaining cases, however, one must invoke a UV completion that allows for mixing with the SM particles or non-renormalizable operators. 
We discuss many such possibilities in Section 2.1. However, in order to assess in full generality the present viability of the Higgs boson and $Z$-boson portals, 
here we treat all DM couplings as free parameters within a simple model-independent bottom-up approach. 
As in previous such analyses \see{Escudero:2016gzx,Arcadi:2017kky}, we disregard UV-completion-dependent 
details concerning the origins of any higher-order operators and assume that any new states in the UV-completions are so massive 
that they cannot impact the phenomenology. This approach allows us to discuss the phenomenology of all Higgs- and $Z$-boson portal scenarios 
on the same footing, and captures the essence of more involved UV constructions that recover the proposed setup in generic regions of their parameter spaces.

Despite their simplicity, SM-portal models exhibit a rich phenomenology, with potential signals in
direct and indirect detection (ID) experiments,
 as well as in collider searches for DM. The absence of such signals 
result in important constraints on their parameters, which were studied for scalar DM with a Higgs portal
in~\cite{Cuoco:2016jqt,He:2016mls,Athron:2017kgt,Bell:2017rgi}, for Higgs portals with DM of spin 0, 1/2 and 1 in~\cite{Beniwal:2015sdl,Chang:2017gla}, 
and for Higgs and $Z$ portals with DM of spin 0, 1/2 and 1 in~\cite{Escudero:2016gzx}. The null results of the many
searches have led to the suspicion that the plausibility of simple WIMP Higgs and $Z$-portal models has been severely
damaged~\see{Arcadi:2017kky, Baum:2017kfa, Casas:2017jjg}. However, neither this nor the impact of planned DD experiments
such as LZ~\cite{Mount:2017qzi} and XENONnT~\cite{Aprile:2015uzo} has yet been quantified by a dedicated statistical analysis. 

We address this point in this work by performing statistical analyses of the spin 0, 1/2 and 1 SM-portal models described in \refsec{Sec:models}.  We use frequentist and Bayesian methodologies for the first time jointly assessing the plausibility of these models.  Our statistical framework is recapitulated in \refapp{Sec:stats}, and we add a novel determination the number of constrained degrees of freedom at the end of \refsec{Sec:results}.
Our work includes data from DD, ID and 
collider experiments, and we also analyze the potential impacts of future DD experiments, using the pseudodata described in \refsec{Sec:like}. 
Our priors for the model parameters are presented in \refsec{Sec:Priors}.
Finally, we present our findings on the viability of SM-portal models in \refsec{Sec:results} and conclude in \refsec{Sec:conclusions}.

Unlike previous analyses~\see{Escudero:2016gzx, Arcadi:2017kky, Baum:2017kfa, Casas:2017jjg}, which were ambivalent about the status of portal models, we find both Higgs- and $Z$-portal models
that are entirely consistent with the available experimental constraints. These include Higgs-portal models
with real scalar, real vector, Majorana or Dirac fermion dark matter particles, as well as
Z-portal models with Majorana or Dirac fermion dark matter particles. Many of the viable models feature relatively light
dark matter particles that would contribute to invisible Higgs or $Z$ decays, and there are also interesting possibilities
for heavier dark matter particles that weigh $\gtrsim 100$~GeV. The planned direct dark matter search experiments
would be sensitive to some of these dark matter models, but Higgs- and $Z$-portal models with Majorana or Dirac 
fermion dark matter particles could still escape detection. 
We ascribe the difference in emphasis between our paper and other analyses to the fact that we calculate statistical measures of goodness-of-fit and relative plausibility of models; some previous analyses drew conclusions on the status of portal models without a clear statistical methodology.
Our results open the way towards constructing viable DM models that are not significantly damaged by the present data. As with other simplified models, the models we discuss do not attempt to capture details and potential issues within any specific UV-complete model. Although we discuss several viable UV completions in Section 2.1 in this simplified model approach, we do not include theoretical or experimental constraints upon the parameter spaces of any particular UV completions. The merit of this approach, which has proven extremely popular in DM \see{Beniwal:2015sdl,Chang:2017gla,Cuoco:2016jqt,He:2016mls,Athron:2017kgt,Bell:2017rgi,Escudero:2016gzx,Berlin:2014tja,Buckley:2014fba,Buchmueller:2014yoa,DiFranzo:2013vra,Harris:2014hga} and LHC studies \see{Abercrombie:2015wmb,Abdallah:2015ter,Alves:2011wf,Boveia:2016mrp,Kraml:2013mwa,Choudhury:2015lha,Chatrchyan:2013sza,Cohen:2013xda}, is that we may study the phenomenology of a comprehensive set of SM portal models capturing common features.

\section{Models}\label{Sec:models}



We consider simplified WIMP models in which spin 0, spin 1/2 or vector DM particles are stabilized 
by a $\mathbb{Z}_2$ symmetry \see{McDonald:1993ex} and couple to the SM Higgs or $Z$ boson. 
This portal interaction with the SM has a number of phenomenological signatures. Through such portals, 
DM can scatter with quarks via $t$-channel exchanges producing signatures that are in principle observable at DD experiments. 
Portal couplings are also responsible for the annihilations of DM particles into SM particles, 
which control the relic abundance of DM via the freeze-out mechanism. The same DM annihilations
could, furthermore, still be active in regions of space characterized by a high DM density, 
such as the galactic center, dwarf spheroidal galaxies (dSph) and possibly the Sun. Primary or secondary traces of annihilation 
could then be detected by ID experiments. In order to explore the vast phenomenology of these simplified models, 
we implemented them in \texttt{FeynRules-2.3.27}\cite{Christensen:2008py,Christensen:2009jx}, 
linked with \mo via \texttt{calcHEP}\cite{Belyaev:2012qa}.

\subsection{The Higgs Portal}
    
We consider models of spin 0, 1/2 and 1 DM particles coupling to the SM via a Higgs portal 
(see e.g.,~\refcite{Arcadi:2017kky,Escudero:2016gzx} for similar Higgs portal models), assuming that
the DM is stabilized by a $\mathbb{Z}_2$ or $U(1)$ symmetry \see{Arcadi:2017kky}. Furthermore, 
in order to accommodate interactions between the Higgs and pairs of photons or gluons via loops, 
we include effective operators $h F_{\mu\nu}F^{\mu\nu}$ that result from the top quark and the $W$-boson loops in the SM. 
    
\begin{description}
    
    \item[Dirac/Majorana fermion]
        DM particles interact with the SM via
        \begin{equation}
        \LG \supset 
        c\,
        \bar\rchi  
        \left(g_s + i g_p \gamma^5 \right)
        \rchi \, h,
        \end{equation}
        where $g_s$ and $g_p$ are scalar and pseudoscalar couplings, respectively, and $c = 1\,(1/2)$ for the Dirac (Majorana) case.
        We note that such interactions could originate from mixing after electroweak symmetry breaking (EWSB) between an 
        SM singlet, $S$, and the Higgs in a gauge-invariant renormalizable model, as discussed in~\cite{Kim:2008pp}:
        \begin{equation}
        \bar\rchi  
        \left(g_s + i g_p \gamma^5 \right)
        \rchi S \to
        \bar\rchi  
        \left(g_s + i g_p \gamma^5 \right)
        \rchi \sin\alpha \, h + \ldots
        \end{equation}
        where $\alpha$ is a mixing angle between the Higgs and singlet. We assume that the singlet is so heavy that it cannot impact the phenomenology. Unitarity would require that\cite{Englert:2011yb}
        \begin{equation}
        \cos^2\alpha \, m_h^2 + \sin^2\alpha \, m_S^2 \lesssim (700\gev)^2,   
        \end{equation}
        where $m_h \simeq 125\gev$ and $m_S$ is the mass of the singlet.
    \item[Scalar]
        DM particles interact with the SM via \see{Burgess:2000yq}
        \begin{equation}
        \LG \supset 
        c\, \lambda  
        \left(v h\modu{\rchi}^2 + \frac{1}{2} h^2 \modu{\rchi}^2 \right),
        \end{equation}
        where $c = 1/2\, (1)$ for a real (complex) scalar and $v=246\gev$ is the vacuum expectation value of the SM Higgs. 
        These interactions could result from a $|\rchi|^2 |H|^2$ operator after EWSB. Unitarity requires that $\lambda \le 8 \pi$\cite{Cynolter:2004cq}.
    \item[Vector]
        DM interacts with the SM via
        \begin{equation}
        \LG \supset 
        c\, g  
        \left(v h \rchi^\mu \rchi_\mu ^\dg + \frac{1}{2} h^2 \rchi^\mu \rchi_\mu ^\dg \right),
        \end{equation}
        where $c = 1/2\,(1)$ for a real (complex) vector and $v=246\gev$ is the vacuum expectation value of the SM Higgs doublet.
        The vector DM may acquire a mass via the Stueckelberg mechanism \see{Ruegg:2003ps}. This interaction could arise by charging the SM Higgs under a new
        $U(1)$ gauge group, which would risk unacceptable mixing between the DM and SM $Z$ boson\cite{DiFranzo:2015nli}, 
        or by an alternative mechanism \see{Duch:2015jta,Lebedev:2011iq,Hambye:2008bq}, e.g., mixing between the SM Higgs
        and a SM singlet scalar that is charged under the $U(1)$.

\end{description}
        
    All of these Higgs portal models allow the possibility that there are no phenomenological signatures of the UV completions in current experiments,
    as we assume here.

\subsection{The $Z$-Boson Portal}

We also consider all possible spin 0, 1/2 and 1  DM particles annihilating via a $Z$ portal. 
    
    \begin{description}
    
    
    \item [Dirac fermion] 
        DM interacts with the SM via
        \begin{equation}\label{Eq:dirac_z}
        \LG \supset \bar\rchi \gamma^\mu 
        \left(g_v + g_a \gamma^5 \right)
        \rchi \, Z_\mu,
        \end{equation}
        where $g_v$ and $g_a$ are vector and axial couplings, respectively. This operator could originate after EWSB by coupling the fermion to the current\cite{Kearney:2016rng,deSimone:2014pda}
        \begin{equation}\label{eq:Z_eff}
        H^\dagger D_\mu H \to \frac12 i g v^2 Z_\mu + \ldots
        \end{equation}
        via a non-renormalizable operator, kinetic mixing between the SM $Z$ boson and a $Z^\prime$ boson~\cite{Arcadi:2014lta}
         or if the DM has an $SU(2) \times U(1)$ charge~\cite{Nagata:2014aoa}. The non-renormalizable operator could, however, 
         generate DM annihlation into $hh$ and $Zh$. Assuming that it originates from a non-renormalizable operator, the interaction in \refeq{Eq:dirac_z} is unitary for couplings and masses of phenomenological interest~\cite{Arcadi:2014lta}.
    
    \item[Majorana fermion]
        DM interacts with the SM via
        \begin{equation}
        \LG \supset \frac{g_a}{2}   
        \bar\rchi \gamma^\mu \gamma^5 \rchi \, Z_\mu.
        \end{equation}
        The vector coupling is forbidden by charge conjugation for Majorana fermions as the operator is odd. 

    \item[Scalar]
        DM interacts with the SM via
        \begin{equation}
        \LG \supset  ig\, \rchi^{\dg} \pdarrow 
        \rchi \, Z_{\mu} + g^2 \modu{\rchi}^2 Z^\mu Z_\mu,
        \end{equation}
        where $a\,\pdarrow b \coloneqq a (\pd^\mu b) - b (\pd^\mu a)$ and $g$ is an effective gauge coupling. 
        These operators may originate from kinetic mixing between the SM $Z$ boson and a $Z^\prime$ boson~\cite{Arcadi:2014lta}, or via higher-order terms of the form
        \begin{equation}
            \LG\supset g' \, (\rchi^{\dg} \pdarrow \rchi) \frac{H^{\dg}\accentset{\leftrightarrow}{D}_{\mu} 
        H}{\mu^2}\,.
        \end{equation}
    
    \item[Vector]
        DM interacts with the SM via
        \begin{equation}\label{Eq:VectorZ}
        \LG \supset  ig \,
        \left( 
        Z^\mu \rchi^{\nu\dagger}\,\pd_{[\mu}\rchi_{\nu]}
        +
        \rchi^\dagger_\mu \rchi_\nu \, \pd^\mu Z^\nu
        \right)
        + \hc,
        \end{equation}
        where $g$ is an effective gauge coupling \see{Arcadi:2017kky}. The higher-order operators that can induce these interactions are rather involved; we refer the reader to Ref.s~\cite{Anastasopoulos:2006cz,Antoniadis:2009ze,Dudas:2009uq,Dudas:2013sia} for possible realizations. 
        As in the Higgs-portal case, we assume here that the vector DM acquires a mass via the Stueckelberg mechanism. 
    \end{description}

We note that the couplings of the $Z$ boson to particles in specific representations of the SM gauge group must take discrete
values in terms of the SU(2) and U(1) gauge couplings. However, more general values are allowed in the presence of mixing
between the DM and SM particles. In the interest of generality, we allow for this possibility without entering into specific mixing
scenarios. We note also that in the spin 1/2 and 1 cases there are unitarity issues (anomaly cancellation and rapid increase
in scattering amplitudes at high energies, respectively) that need to be addressed in the UV-completion \see{Kearney:2016rng}, whose possible
experimental signatures we do not discuss here.

\section{Likelihood}\label{Sec:like}

The likelihood function is a critical ingredient for our frequentist and Bayesian statistical methodologies, which are described in \refapp{Sec:stats}. 
Our likelihood, summarized in \reftable{Table:Data}, includes collider, ID and DD searches for DM, as well as the 
Planck determination of the DM relic abundance. Our implementations of the likelihood contributions required are detailed in the forthcoming Sections. As explained in \refsec{Sec:DD} and \refsec{Sec:ID} we approximate experimental limits on DD and ID cross sections by step functions, similar to the treatment in e.g., \refcite{Arcadi:2017kky}. We, however, incorporate important nuclear and astrophysical uncertainties.
\vspace{0.5cm}

\subsection{Relic density}

We assume that the DM candidate accounts of all of the observed DM, i.e., it is not an underabundant species of DM. We calculate the relic abundance of DM with \mo. We describe the Planck determination of the relic density\cite{Ade:2015xua} 
by a Gaussian, and include in quadrature a supplementary $10\%$ theoretical uncertainty in the calculation of the relic abundance. 
Recent analyses suggest that the standard treatment in \mo may be insufficient, as it neglects the $s$-dependence of the mediator width in a Breit-Wigner and the effect of kinetic decoupling\cite{Binder:2017rgn,Duch:2017nbe}, but we leave any improvement to future work. 


\subsection{Collider constraints}

\subsubsection{Invisible decay widths}

The invisible width of the $Z$-boson was measured by LEP\cite{Olive:2016xmw}. We calculate the SM contribution, 
due to $\Gamma(Z\to\nu\nu)$, with a parametric two-loop formula from \refcite{Freitas:2014hra}, which includes the
dependence on $M_Z$, which we treat as a nuisance parameter. We calculate $\Gamma(Z\to\text{DM})$ with \mo and 
implement the measurement as a Gaussian upon $\Gamma(Z\to\nu\nu) + \Gamma(Z\to\text{DM})$. 
The branching fraction of an SM-like Higgs to invisible particles was constrained by CMS\cite{CMS-PAS-HIG-16-016} to be 
$\BR(h \to \text{invisible}) \lesssim 0.24$. Since the likelihood function has been published, we apply it directly in our analysis. 
We neglect any correlations between measurements of the SM masses and widths and we neglect $\BR(h \to \nu\nu)$,
which is absent in the SM.

\subsubsection{Monojet and monophoton searches}\label{sec:LHC}

\begin{figure}
    \centering
    \includegraphics[width=0.7\textwidth]{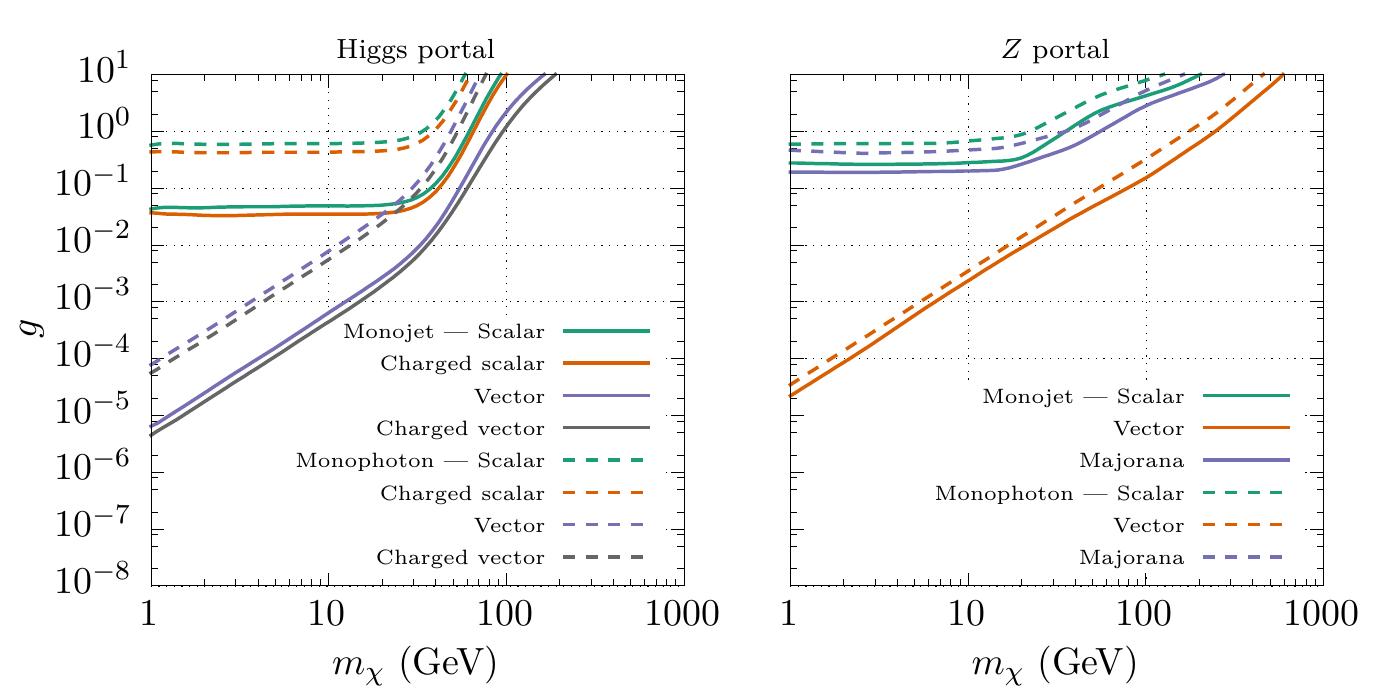}
    \caption{\it Exclusion limits from LHC monojet (solid lines) and monophoton (dashed lines) searches for DM in the H- (left) and $Z$-portal (right) models,
    which have a single coupling $g$ to the SM.}
    \label{Fig:LHC}
\end{figure}

Monojet and monophoton searches look for missing transverse momentum resulting from DM particles produced in 
pairs that escape the detector. The DM particles are not produced back-to-back in the laboratory frame when they
recoil from the emission of a jet or photon. We include $\sqrt{s}=8\tev$ and $13\tev$ monojet and monophoton searches for DM at the LHC via 
\texttt{Check\allowbreak{}MATE-2}\cite{Dercks:2016npn,deFavereau:2013fsa,Cacciari:2011ma,Cacciari:2005hq,Read:2002hq,Cacciari:2008gp}.  
The searches that we included are listed in \reftable{Table:LHC}, and the limits we use are shown in \reffig{Fig:LHC}.
In the models under study, production cross sections at $\sqrt{s} = 8\tev$ and $13\tev$ are comparable, and the greater integrated luminosity in the former make them more powerful~\footnote{We did not include a recent ATLAS monojet search at $\sqrt{s} = 13\tev$ with $36\,\text{fb}^{-1}$\cite{Aaboud:2017phn} that
is not yet implemented in \texttt{Check\allowbreak{}MATE-2}. We do not anticipate that this would change significantly the results we find using the monojet searches that are implemented in \texttt{Check\allowbreak{}MATE-2}.}. Since we add to the SM only a DM particle, 
there are no cascade decays of dark sector particles resulting in \eg jets or leptons.

We generated events from our models with \texttt{MadGraph5\_\allowbreak{}aMC@NLO}\cite{Alwall:2014hca} 
and processed the resulting \texttt{lhe} event files with \texttt{Check\allowbreak{}MATE-2}. We derived our own 
$95\%$ exclusion contours on planes of mass versus coupling for each SM-portal model for the ranges of 
mass and coupling in \reftable{Table:Priors}. We approximated the likelihood function by a step function by assigning
to each parameter point a likelihood of zero if it was excluded at $95\%$ and one otherwise. 

\begin{table}
\centering
\begin{tabular}{lcccc}  
\toprule
Analysis & $\sqrt{s}$ (TeV) & $\int \mathcal{L}$ ($\text{fb}^{-1}$) & \texttt{Check\allowbreak{}MATE-2}\\
\midrule
ATLAS monojet\cite{Aad:2014nra} & $8$ & $20.3$ & \texttt{atlas\_1407\_0608}\\
ATLAS monojet\cite{Aad:2015zva} & $8$ & $20.3$ & \texttt{atlas\_1502\_01518}\\
ATLAS monojet\cite{Aaboud:2016tnv}& $13$ & $3.2$ & \texttt{atlas\_1604\_07773}\\
CMS monojet\cite{Khachatryan:2014rra} & $8$ & $19.7$ & \texttt{cms\_1408\_3583}\\
\midrule
ATLAS monophoton\cite{Aad:2014tda} & $8$ & $20.3$ & \texttt{atlas\_1411\_1559}\\
ATLAS monophoton\cite{Aaboud:2016uro} & $13$ & $3.2$ & \texttt{atlas\_1604\_01306}\\
ATLAS monophoton\cite{Aaboud:2017dor}& $13$ & $36.1$ & \texttt{atlas\_1704\_03848}\\
\bottomrule
\end{tabular}
\caption{\it Monojet and monophoton LHC searches for DM included in our analysis via \texttt{Check\allowbreak{}MATE-2}.}\label{Table:LHC}
\end{table}

\subsection{Direct detection}\label{Sec:DD}


We include the world-leading SI DD constraints from PandaX~\cite{Cui:2017nnn} and world-leading SD DD constraints for 
neutrons from PICO~\cite{Amole:2017dex} and for protons from PandaX~\cite{Fu:2016ega}. 
We have calculated scattering cross sections with \mo while noting, however, that there are appreciable uncertainties that we now discuss
and have considered in our analysis.

The WIMP interactions with partons can be described by an effective field theory. For non-relativistic velocities, scalar, vector, 
pseudovector and tensor operators dominate, as other operators are velocity-suppressed. The scalar, vector, pseudovector and tensor operators 
contribute to SI even, SI odd, SD even and SD odd interactions, respectively\cite{Belanger:2008sj}. The WIMP interactions with nucleons, 
which are relevant for DD, are governed by the partonic interactions and nuclear form factors. The vector, pseudovector and tensor form factors 
are well-known from lattice calculations, and because the vector form-factor depends only upon the valence quarks. We set them to their \mo defaults. 

However, there are uncertainties in the scalar form factors, which depend upon
\begin{align}
\sigma_s &\equiv m_s \langle N | s\bar s | N\rangle,\\
\sigma_{\pi N} &\equiv \frac12 (m_u + m_d) \langle N | u\bar u + d\bar d | N\rangle,
\end{align}
as well as the ratios of light-quark masses. 
There is tension between determinations of $\sigma_{\pi N}$: lattice determinations favor about $40\mev$\cite{Shanahan:2016pla},
whereas phenomenological determinations favor about $60\mev$\cite{Alarcon:2011zs,Alarcon:2012nr,Hoferichter:2015dsa,RuizdeElvira:2017stg}. We modify the \mo default of $34\mev$,
picking a prior that is flat between a precise lattice determination of $37.2\mev$\cite{Abdel-Rehim:2016won} and a recent 
phenomenological determination of $58\mev$\cite{RuizdeElvira:2017stg}, with Gaussian tails equal to the experimental uncertainties. 
To check the sensitivity of our results to this treatment we perform extra calculations in which we discard lattice results and use only the 
phenomenological determination. We pick a Gaussian for $\sigma_s$ from a lattice determination\cite{Abdel-Rehim:2016won}~\footnote{An
alternative approach estimates $\sigma_s$ using $\sigma_{\pi N}$ and phenomenological estimates of the SU(3)-breaking contribution
to baryon masses, $\sigma_0$. However, this is subject to considerable uncertainties in $\sigma_0$~\cite{Alarcon:2012nr} as well as $\sigma_{\pi N}$.} and
flat priors between intervals reported by the PDG\cite{Olive:2016xmw} for ratios of light-quark masses. 
Even though a recent analysis~\cite{Hoferichter:2017olk} indicates that uncertainties in the Higgs-nucleon coupling are likely to be overestimated, for our analysis we choose a conservative approach in our treatment of the DD data.


In addition to nuclear uncertainties, there are astrophysical uncertainties that affect DD, such as the velocity profile and local density of DM. 
We include a log-normal uncertainty upon a canonical choice of local density, $\rho = 0.3\gev/\text{cm}^3$. 
Since we assume that the DM accounts of the entire relic abundance, we do not rescale DD cross sections by e.g., $\Omega h^2 / 0.1$.
For all but the real scalar Higgs portal, 
we neglect velocity profile uncertainties. For that model, though, we vary the shape parameters of a truncated Maxwell-Boltzmann velocity distribution 
describing the velocity profile.


In our implementation of the DD likelihood, we picked a step function at the $90\%$ limits for scattering cross section from 
PandaX and PICO  for a particular DM mass and marginalized uncertainties in the local DM density, nuclear form factors and 
ratios of light quark masses. As a crosscheck, in specific cases, we used the likelihood functions implemented at the event level in 
\ddcalc and marginalized the local DM density and velocity distribution. 

To the best of our knowledge, since we investigated the DM velocity profile, this constitutes the most comprehensive treatment of 
DD uncertainties to date. In, for example, a recent GAMBIT analysis of DD in the real scalar $h$-portal\cite{Athron:2017kgt}, 
only the light quark masses, the local density and form-factor uncertainties were treated with nuisance parameters. Note, moreover, 
that our prior of $\sigma_{\pi N} \approx 27 \text{--} 58 \mev$ differs from that in GAMBIT of $58\pm9\mev$, reflecting the tension
between the lattice and phenomenological determinations.

\subsection{Indirect detection}\label{Sec:ID}

We consider the Fermi-LAT limit from several dwarf spheroidal galaxies (dSphs)~\cite{Ackermann:2015zua} on the zero-velocity limit of the DM annihilation cross section. 
As in the case of DD cross sections, we do not rescale ID cross sections by e.g., $\Omega h^2 / 0.1$ as we assume that the DM accounts of the entire relic abundance.
We approximate the likelihood function by step functions at the $95\%$ exclusion contours on the ($m$, $\langle \sigma v \rangle$) planes for 
$u\bar u$, $b\bar b$, $WW$, $e\bar e$, $\mu\bar\mu$ and $\tau\bar\tau$ channels. The astrophysical uncertainties for ID are characterized by 
uncertainties in $J$-factors for each dSph. Since we consider Fermi-LAT's combined limit from several dSphs, we make an approximation: 
we model uncertainties in the limit by a universal $J$-factor uncertainty that scales the whole signal,
picking an uncertainty in $\log_{10} J$ of $0.25$ motivated by typical $J$-factor uncertainties for individual galaxies. Of the 15 dSphs included in the Fermi-LAT limit, the greatest $\log_{10} J$ uncertainty was $0.31$ and the smallest was $0.16$. Since a universal $J$-factor implies that the uncertainties are $100\%$ correlated, we anticipate that our treatment is conservative (i.e., overestimates the uncertainties). We could, in principle, perform a detailed treatment of the Fermi-LAT data with 15 individual $J$-factors. For the portal models under consideration, however, the ID cross sections are typically smaller than the limits from Fermi-LAT and we argue that our simplified treatment is adequate.

We do not include constraints from solar neutrinos from IceCube~\cite{Aartsen:2016zhm}. IceCube may compete with PICO limits 
on the spin-dependent scattering cross section, though this interpretation requires assumptions about DM spin-independent 
interactions in the Sun, the square of the local density, whether DM reaches equilibrium in the Sun, 
and the rates at which DM annihilates to specific final states. We anticipate that a careful treatment of uncertainties, 
as discussed in \refcite{Bagnaschi:2017tru}, would render the IceCube limit weaker than or similar to that from PICO in our models.

\subsection{Future data}

\begin{figure}
    \centering
    \includegraphics[width=0.96\textwidth]{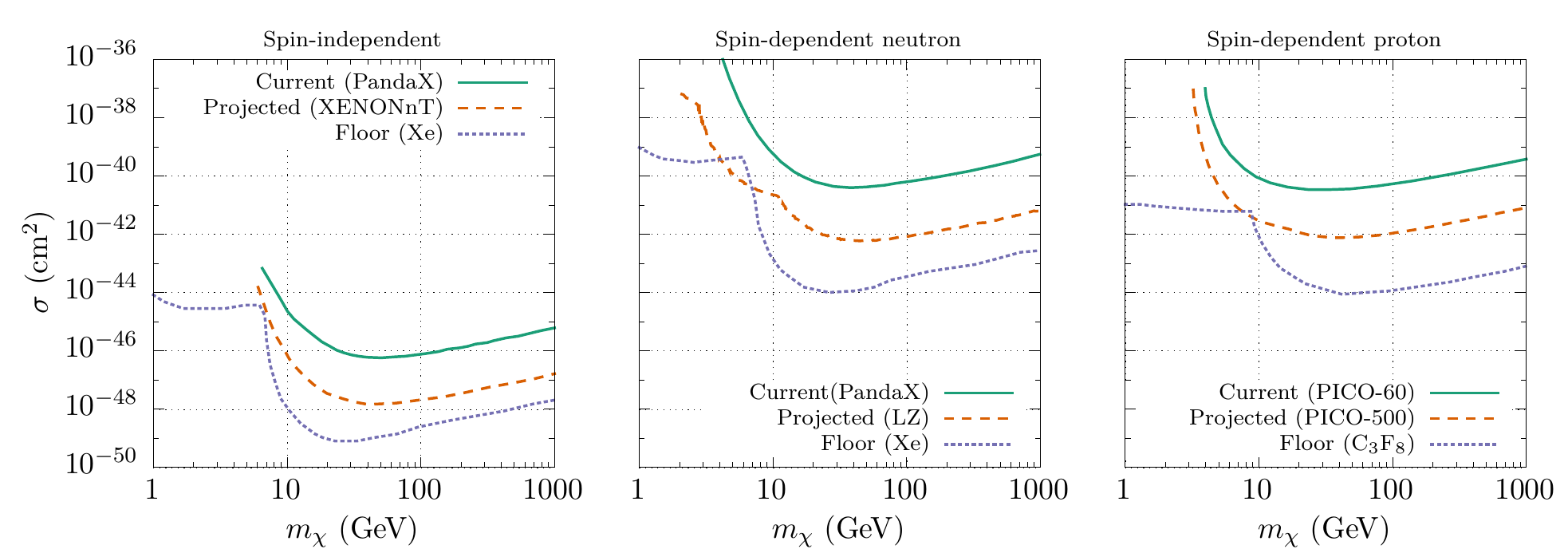}
    \caption{\it Current (green solid lines), projected (orange dashed lines) and neutrino floor (blue dotted lines) $90\%$ limits for DD searches for DM in spin-independent (left), spin-dependent neutron (center) and spin-dependent proton interactions. See \reftable{Table:Data} for further details.}
    \label{Fig:DD}
\end{figure}

In assessing the potential impacts of future experiments,
we utilize projected $90\%$ exclusions from LZ~\cite{Cushman:2013zza}, XENONnT~\cite{Aprile:2015uzo} and PICO500~\cite{pico}, 
and an estimate of the neutrino floor --- the level of scattering cross section at which a WIMP signal becomes hard to 
distinguish from the neutrino background~\cite{Ruppin:2014bra}. The neutrino floors assume Xe targets, 
except for the SD interaction with protons, which assumes a C$_3$F$_8$ target, as in the PICO experiment. 
%
%
The neutrino floor can be very slowly lowered with greater exposure (or overcome with e.g., directional detection, modulation or complementarity between different target nuclei);
see \refcite{Ruppin:2014bra} for details of the assumed exposures. The current limits, projected limits and neutrino floors are shown in \reffig{Fig:DD}.

The projected limits are expected limits assuming the background only hypothesis, i.e., no DM. The observed limit would, however, be subject to statistical fluctuations. We do not investigate the impact of such fluctuations and assume that evidences calculated with projected limits are similar to ones for which the fluctuations are averaged. We anticipate that this is reasonable to within the desired uncertainty of about an order of magnitude in the evidence. 

\section{Priors}\label{Sec:Priors}

The priors for the model parameters used in our Bayesian analysis are shown in \reftable{Table:Priors}. 
In addition to the DM mass and couplings, there are numerous nuisance parameters describing uncertainty about,
\eg the local density of DM, as discussed in \refsec{Sec:like}. Our priors for the DM mass and couplings are quite relaxed; 
we permit masses between $1\gev$ and $10^4\gev$ and couplings between $10^{-6}$ and $4\pi$. Since we are {\it a priori} ignorant of their scale, 
we pick logarithmic priors for the mass and couplings, \eg $\pg{\ln m}{M} = \text{const}$. For models with two couplings, 
we pick factorizable priors, reflecting the assumption that the couplings are determined by different mechanisms and thus independent, 
\eg $\pg{g_s, g_p}{M} = \pg{g_s}{M} \times \pg{g_p}{M}$.
We investigate the sensitivity to priors in \refsec{Sec:sensitivity}. 

\section{Results}\label{Sec:results}

We interfaced our private code to \texttt{(Py-)MultiNest-3.10}\cite{Buchner:2014nha,Feroz:2007kg,Feroz:2008xx,2013arXiv1306.2144F} 
and analyzed our data with \texttt{SuperPlot}\cite{Fowlie:2016hew}. We used $5000$ live points and a stopping tolerance of $10^{-4}$,
and selected importance sampling determinations of evidences and posteriors.

\begin{figure}
    \centering
    \begin{subfigure}[t]{0.49\textwidth}
        \centering
        \includegraphics[width=\textwidth]{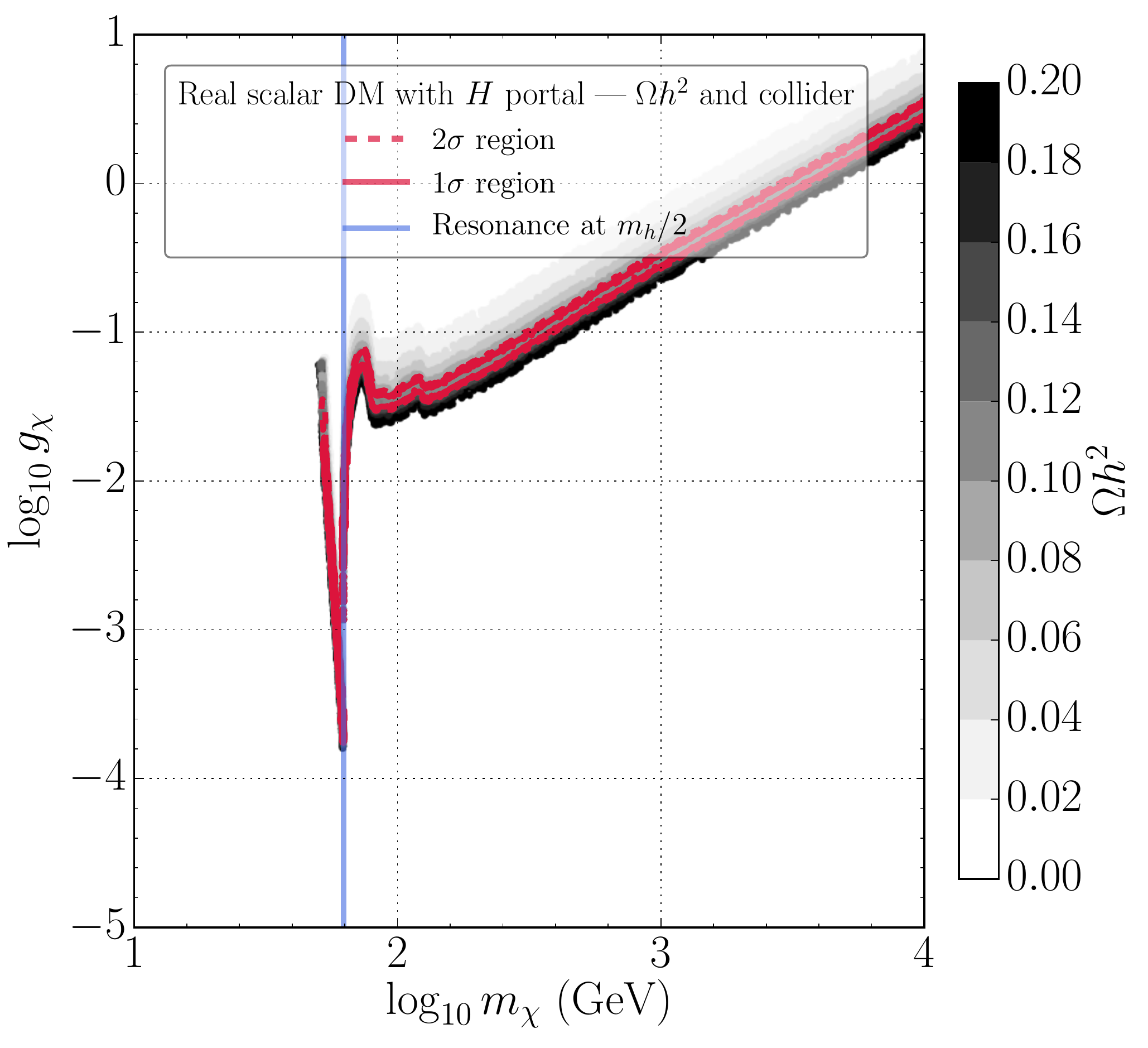}
        \caption{\it Relic density and collider constraints.}
        \label{Fig:}
    \end{subfigure}
    \begin{subfigure}[t]{0.49\textwidth}
        \centering
        \includegraphics[width=\textwidth]{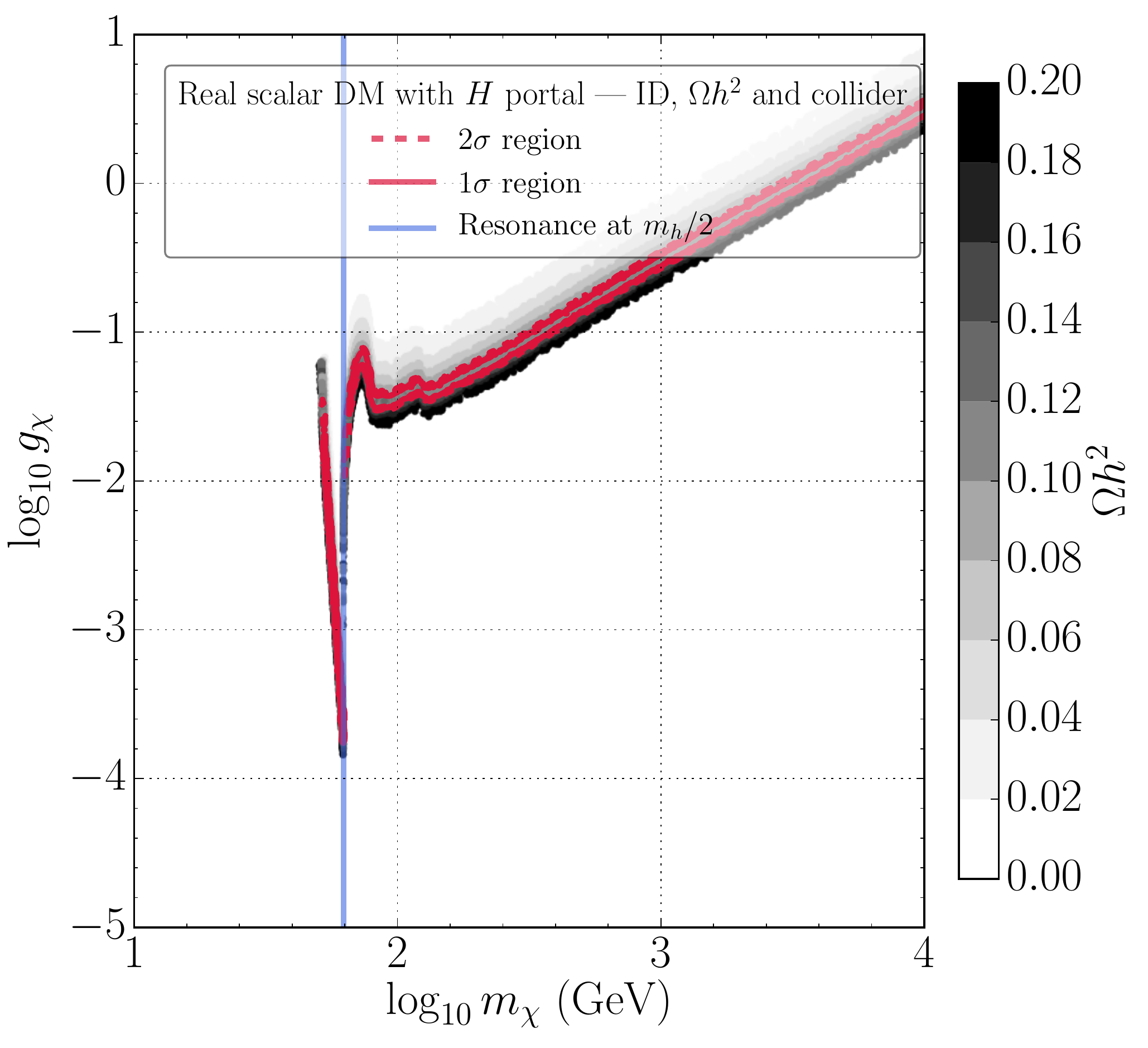}
        \caption{\it Indirect detection, relic density and collider \\ constraints.}
        \label{Fig:}
    \end{subfigure}
    \begin{subfigure}[t]{0.49\textwidth}
        \centering
        \includegraphics[width=\textwidth]{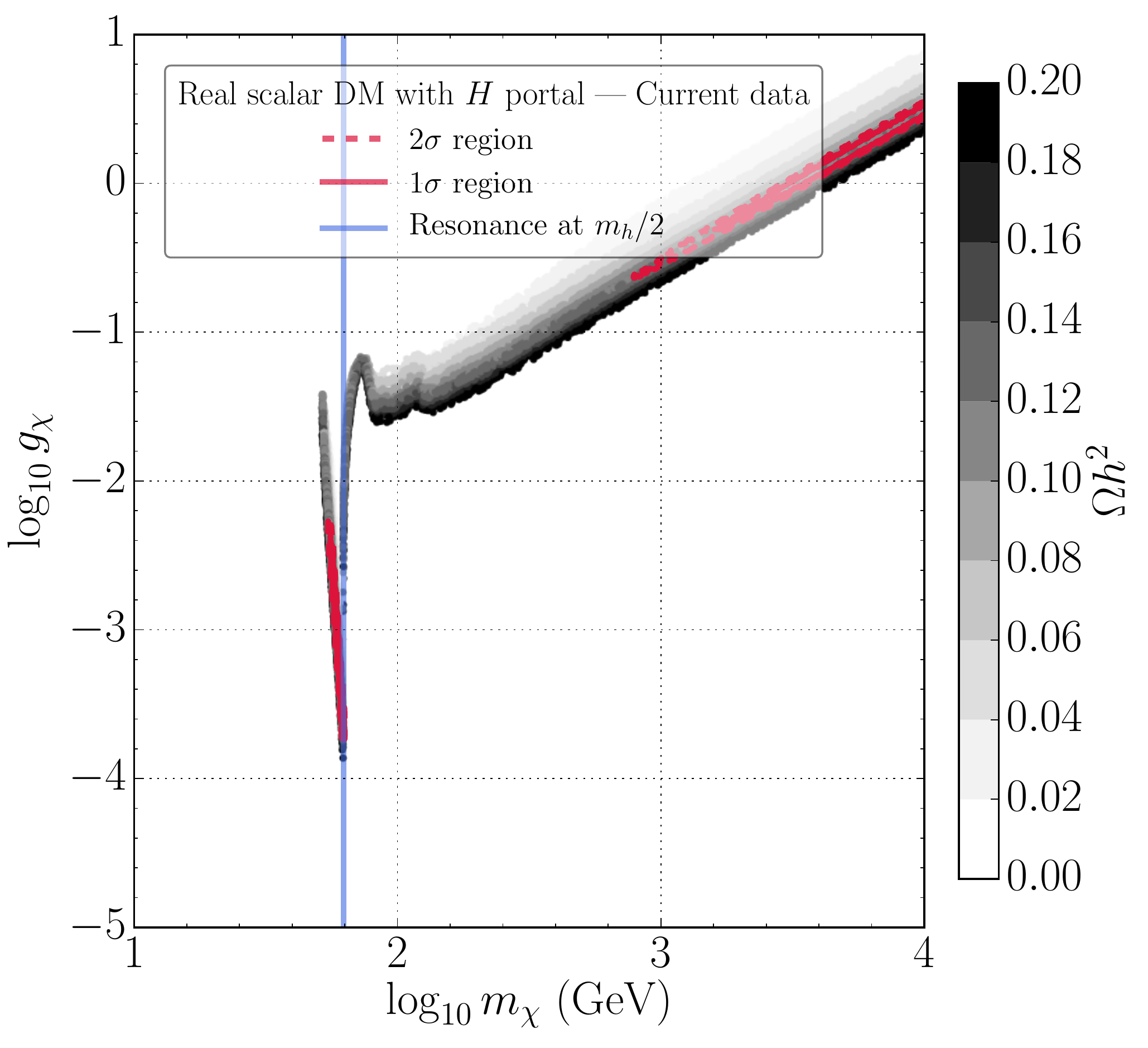}
        \caption{\it Direct and indirect detection, relic density and \\ collider constraints.}
        \label{Fig:}
    \end{subfigure}
    \begin{subfigure}[t]{0.49\textwidth}
        \centering
        \includegraphics[width=\textwidth]{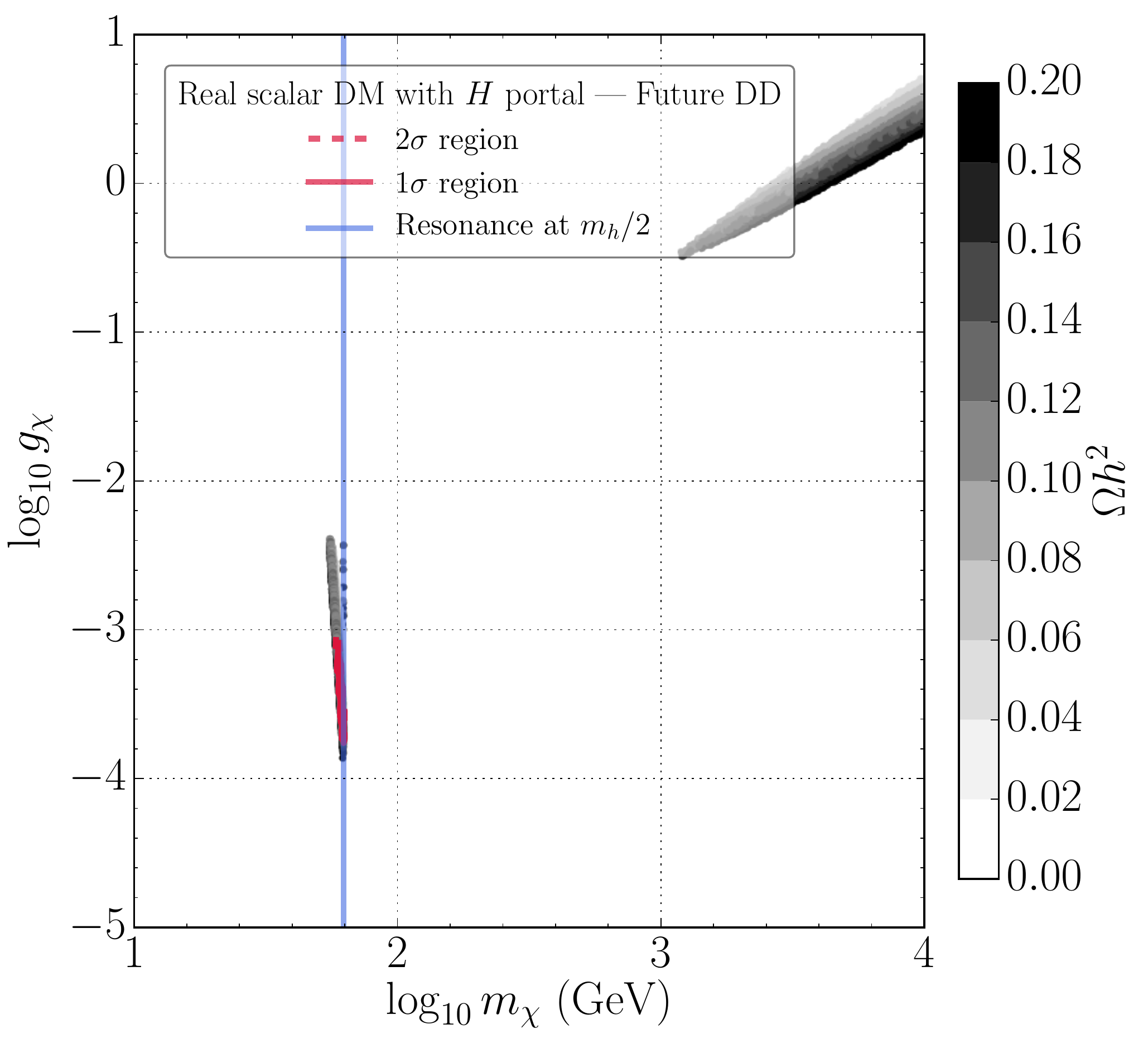}
        \caption{\it Possible LZ constraints on direct detection, indirect detection, relic density and collider constraints.}
        \label{Fig:}
    \end{subfigure}
    \caption{\it Confidence regions of DM mass and coupling for real scalar Higgs-portal DM in light of data from various experiments.}
    \label{Fig:mass_coupling}
\end{figure}

\begin{figure}
    \centering
    \begin{subfigure}[t]{0.49\textwidth}
        \centering
        \includegraphics[width=\textwidth]{dd_id_rel_col__neutral_scalar_h_portal_2_3.pdf}
                \caption{\it Real scalar, all present constraints}
        \label{Fig:}
    \end{subfigure}
    \begin{subfigure}[t]{0.49\textwidth}
        \centering
        \includegraphics[width=\textwidth]{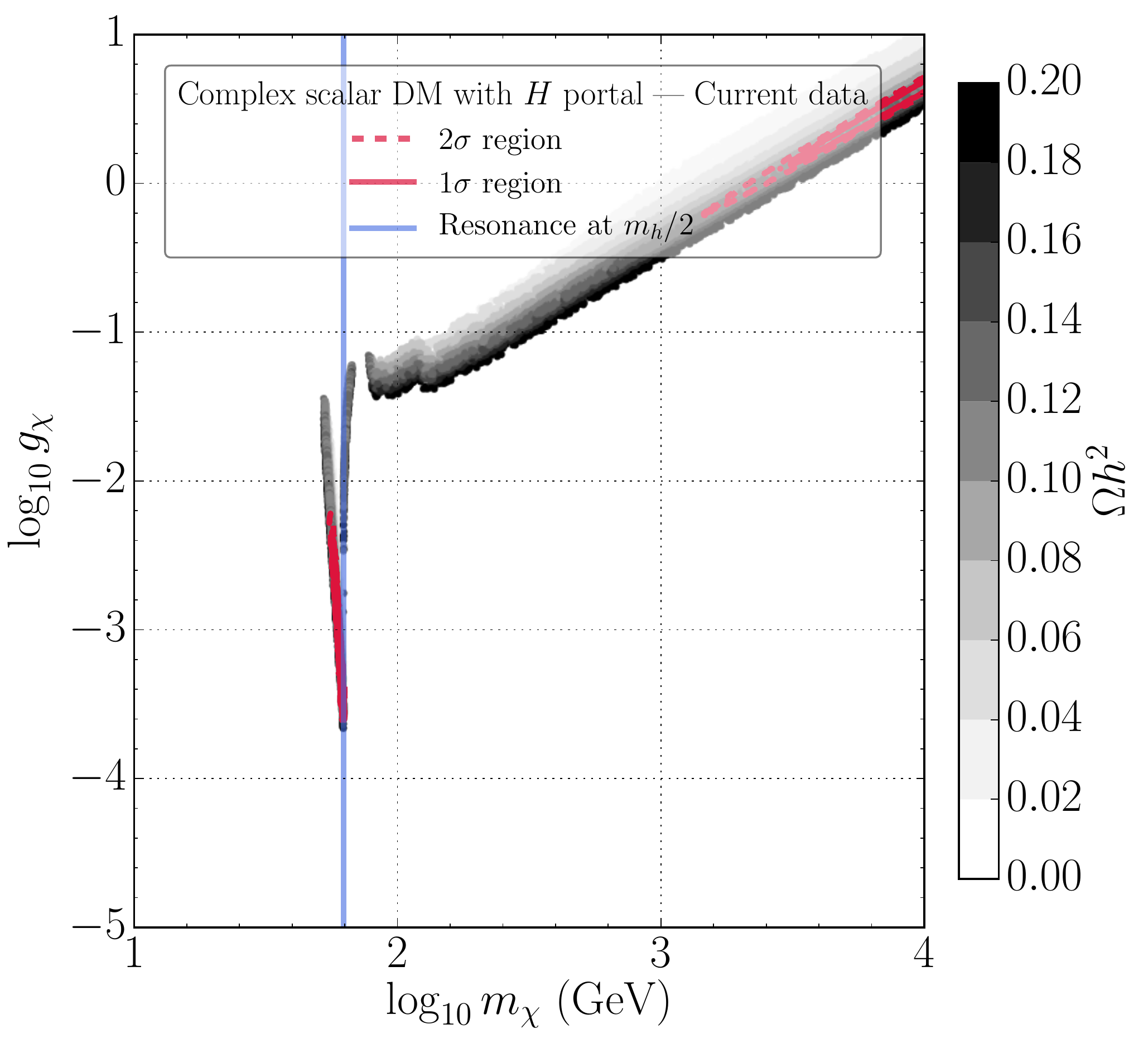}
        \caption{\it Complex scalar, all present constraints}
        \label{Fig:}
    \end{subfigure}
    \begin{subfigure}[t]{0.49\textwidth}
        \centering
        \includegraphics[width=\textwidth]{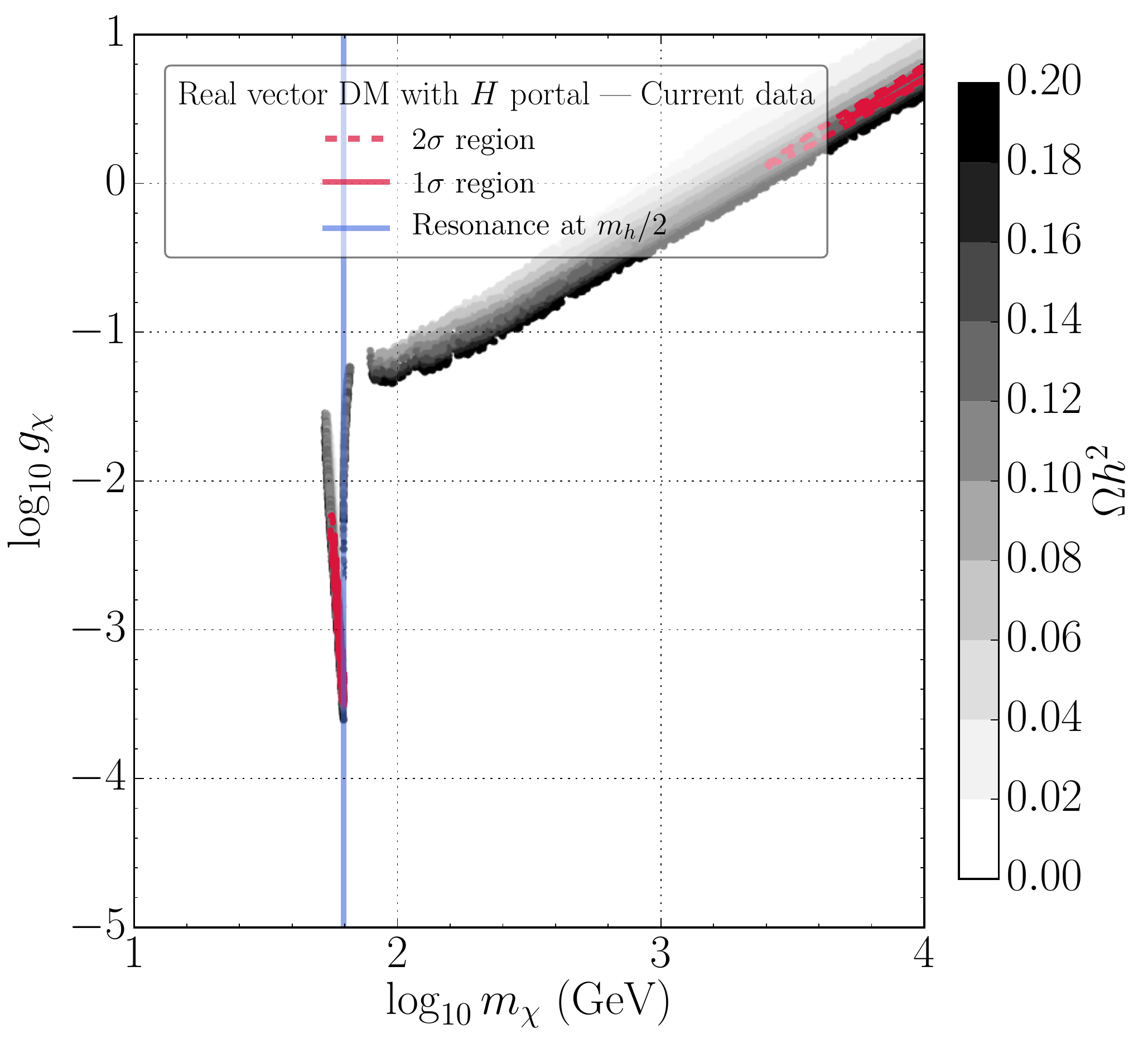}
        \caption{\it Real vector, all present constraints}
        \label{Fig:}
    \end{subfigure}
    \begin{subfigure}[t]{0.49\textwidth}
        \centering
        \includegraphics[width=\textwidth]{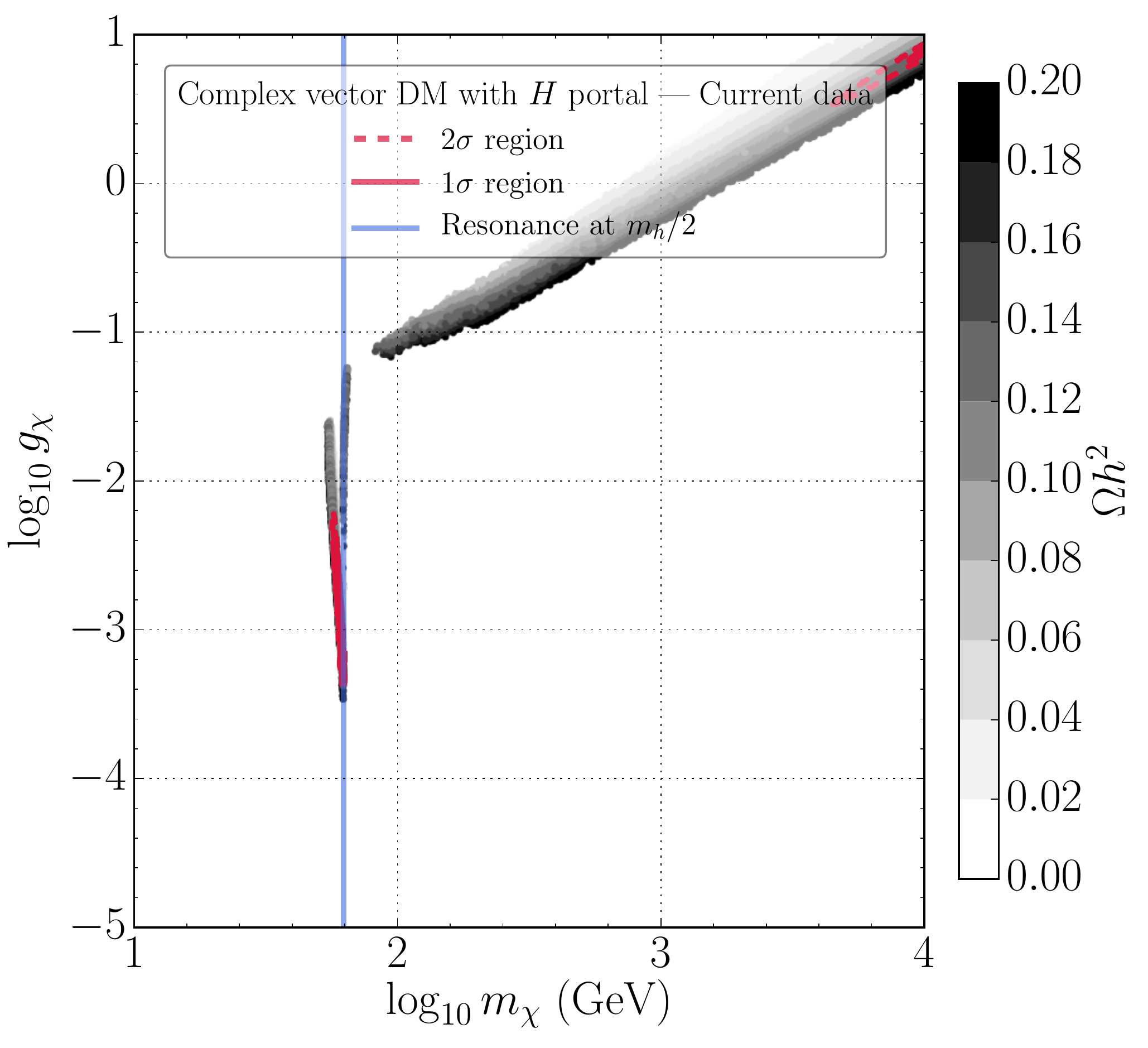}
        \caption{\it Complex vector, all present constraints}
        \label{Fig:}
    \end{subfigure}
    \caption{\it Confidence regions of DM mass and coupling for Higgs-portal scalar and vector DM models in light of all present data.}
    \label{Fig:mass_coupling_vector_scalar_h_portal}
\end{figure} 

\begin{figure}
    \centering
    \begin{subfigure}[t]{0.49\textwidth}
        \centering
        \includegraphics[width=\textwidth]{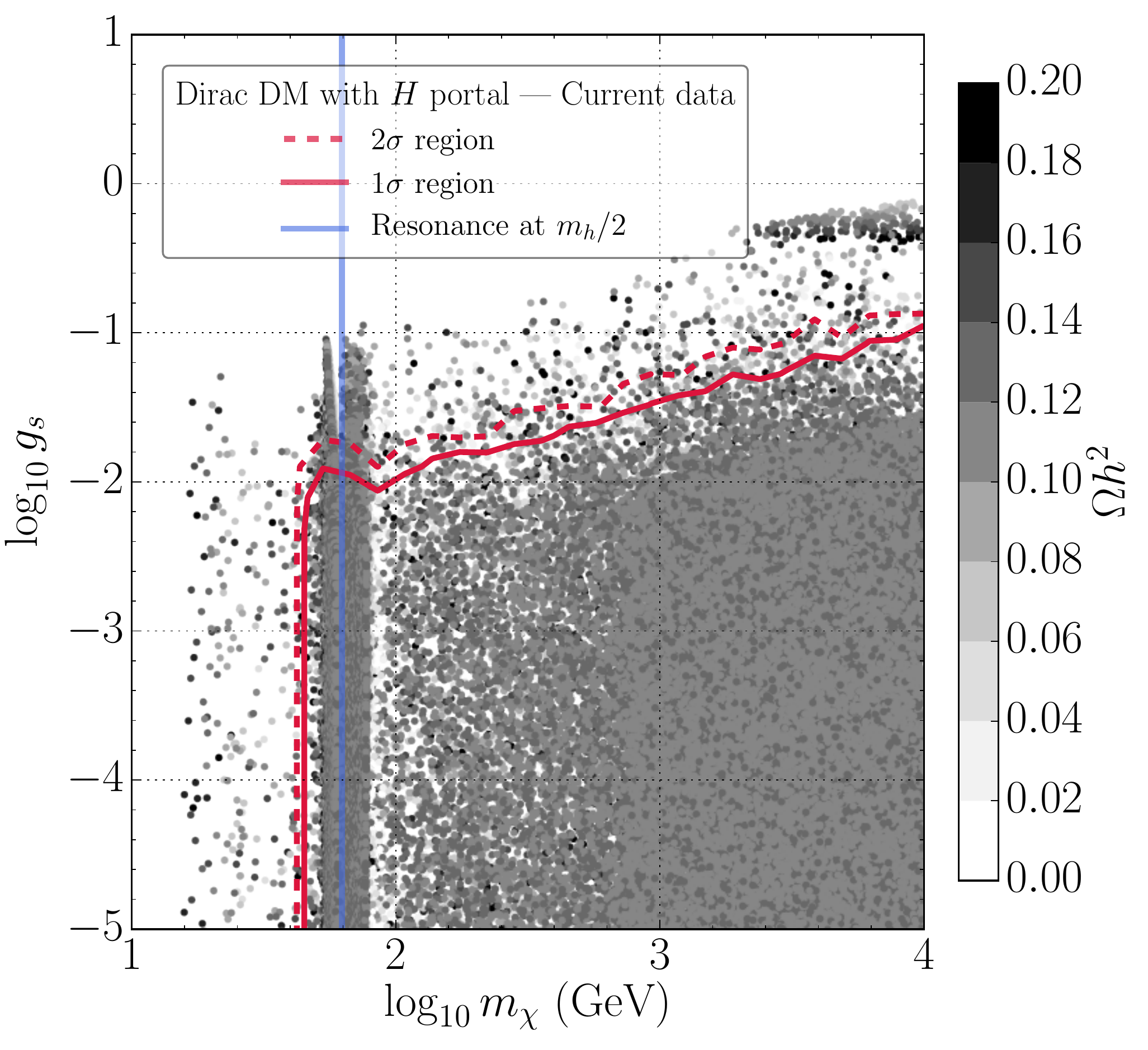}
        \caption{\it Dirac DM, scalar coupling}
        \label{Fig:}
    \end{subfigure}
    \begin{subfigure}[t]{0.49\textwidth}
        \centering
        \includegraphics[width=\textwidth]{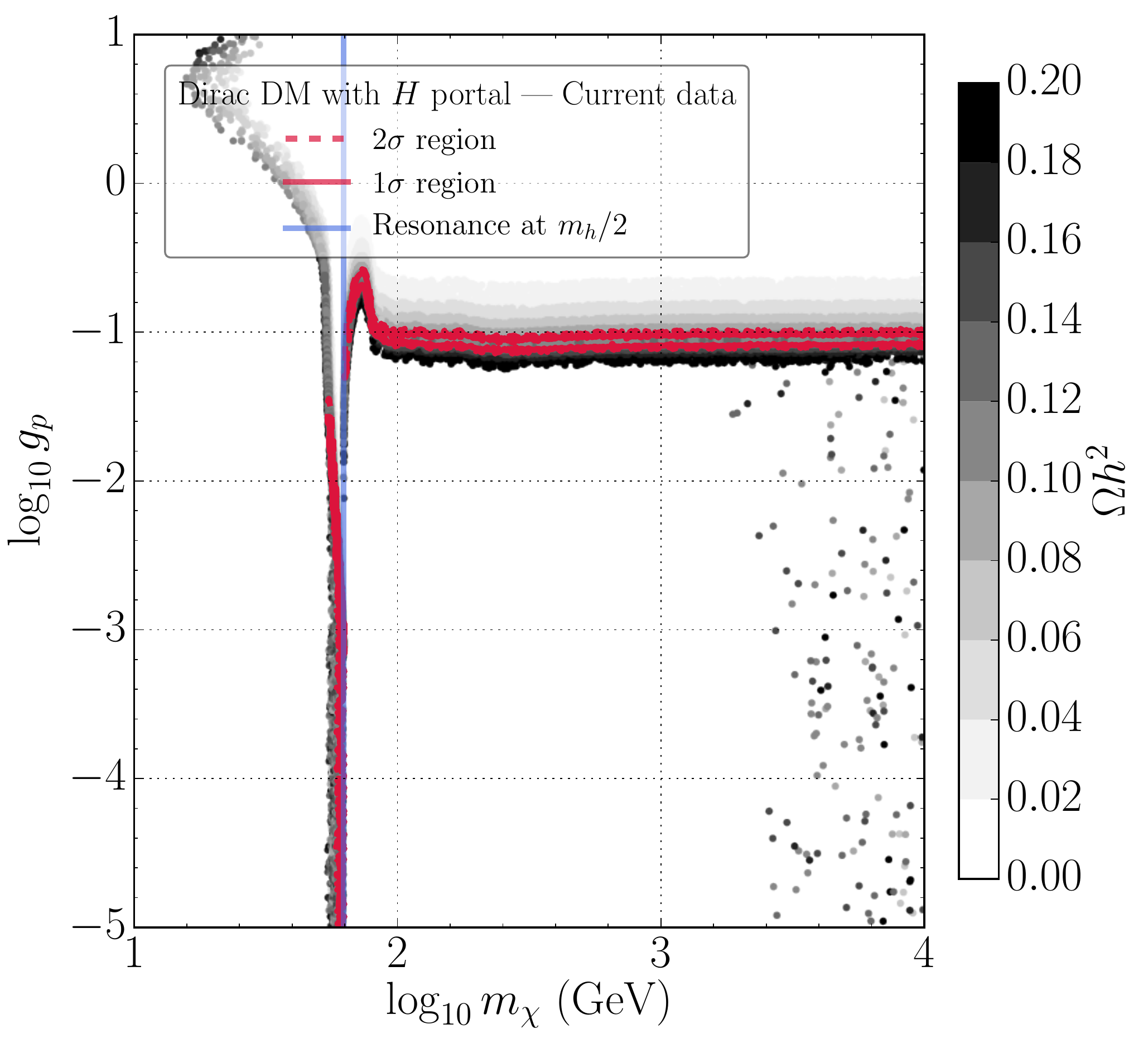}
        \caption{\it Dirac DM, pseudoscalar coupling}
        \label{Fig:}
    \end{subfigure}
    \begin{subfigure}[t]{0.49\textwidth}
        \centering
        \includegraphics[width=\textwidth]{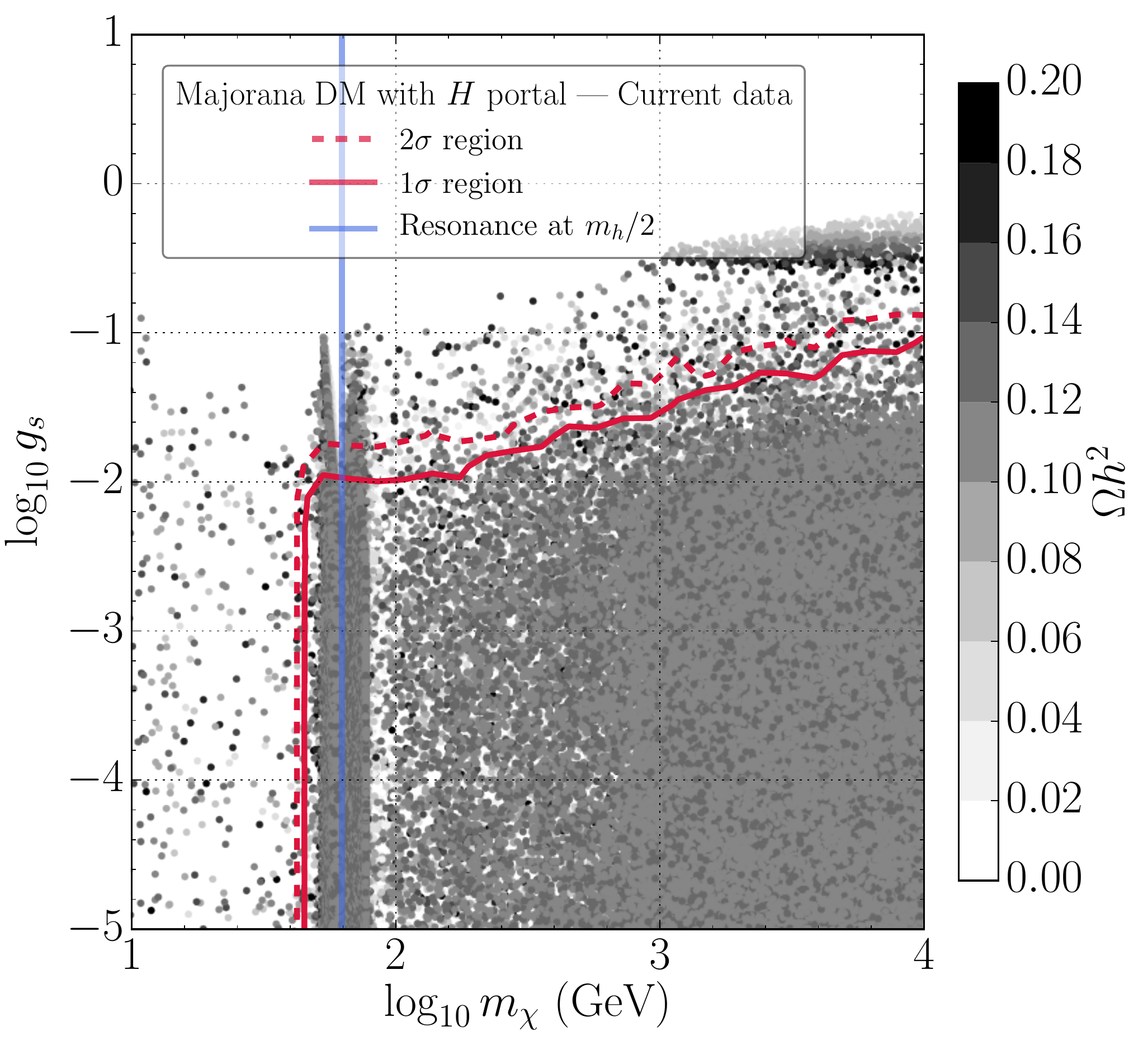}
        \caption{\it Majorana DM, scalar coupling}
        \label{Fig:}
    \end{subfigure}
    \begin{subfigure}[t]{0.49\textwidth}
        \centering
        \includegraphics[width=\textwidth]{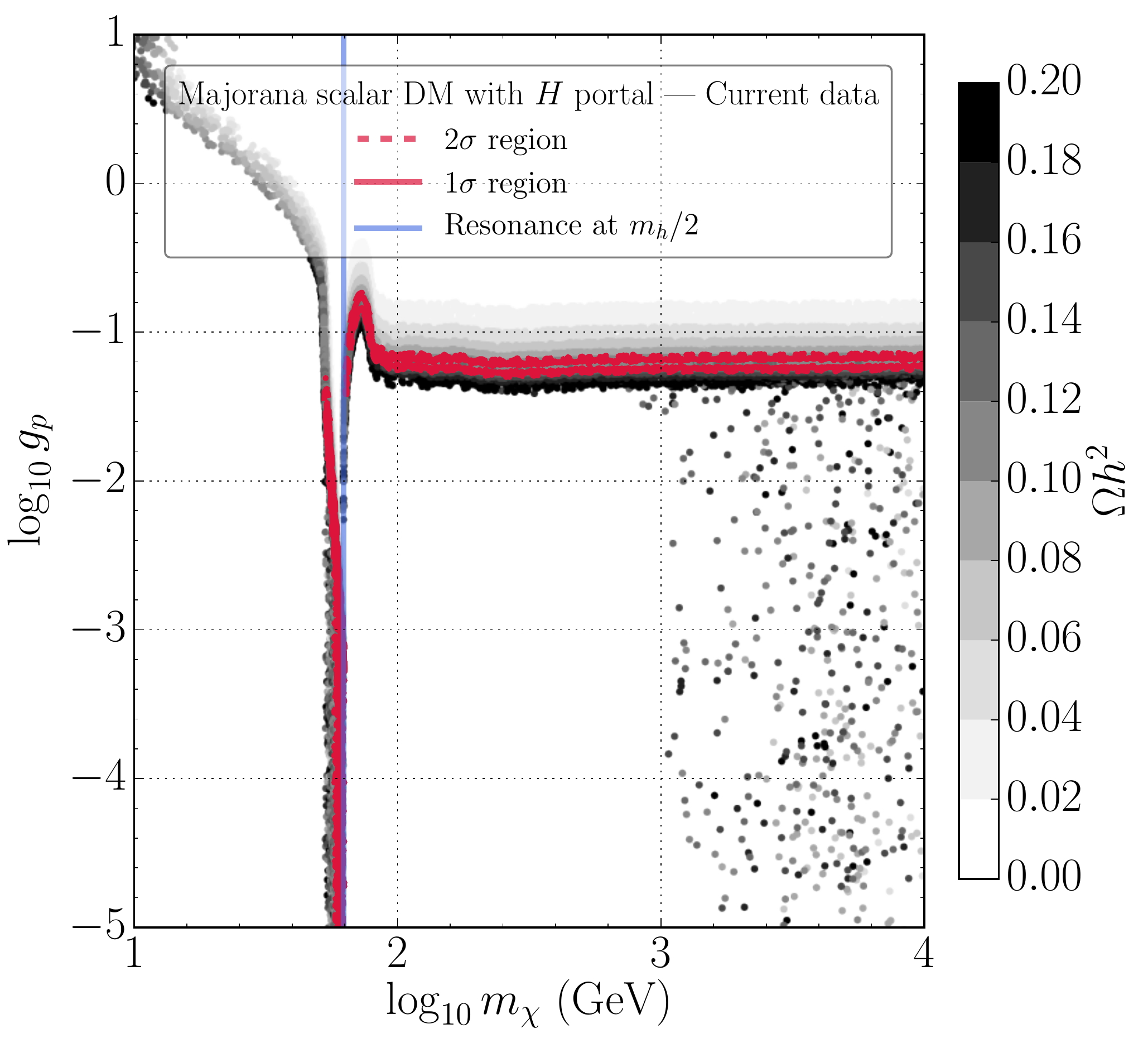}
        \caption{\it Majorana DM, pseudoscalar coupling}
        \label{Fig:}
    \end{subfigure}
    \caption{\it Confidence regions of DM mass and coupling for Higgs-portal fermionic DM models in light of all current data.}
    \label{Fig:mass_coupling_fermion_h_portal}
\end{figure}    
  
\begin{figure}
    \centering
    \begin{subfigure}[c]{0.49\textwidth}
        \centering
        \includegraphics[width=\textwidth]{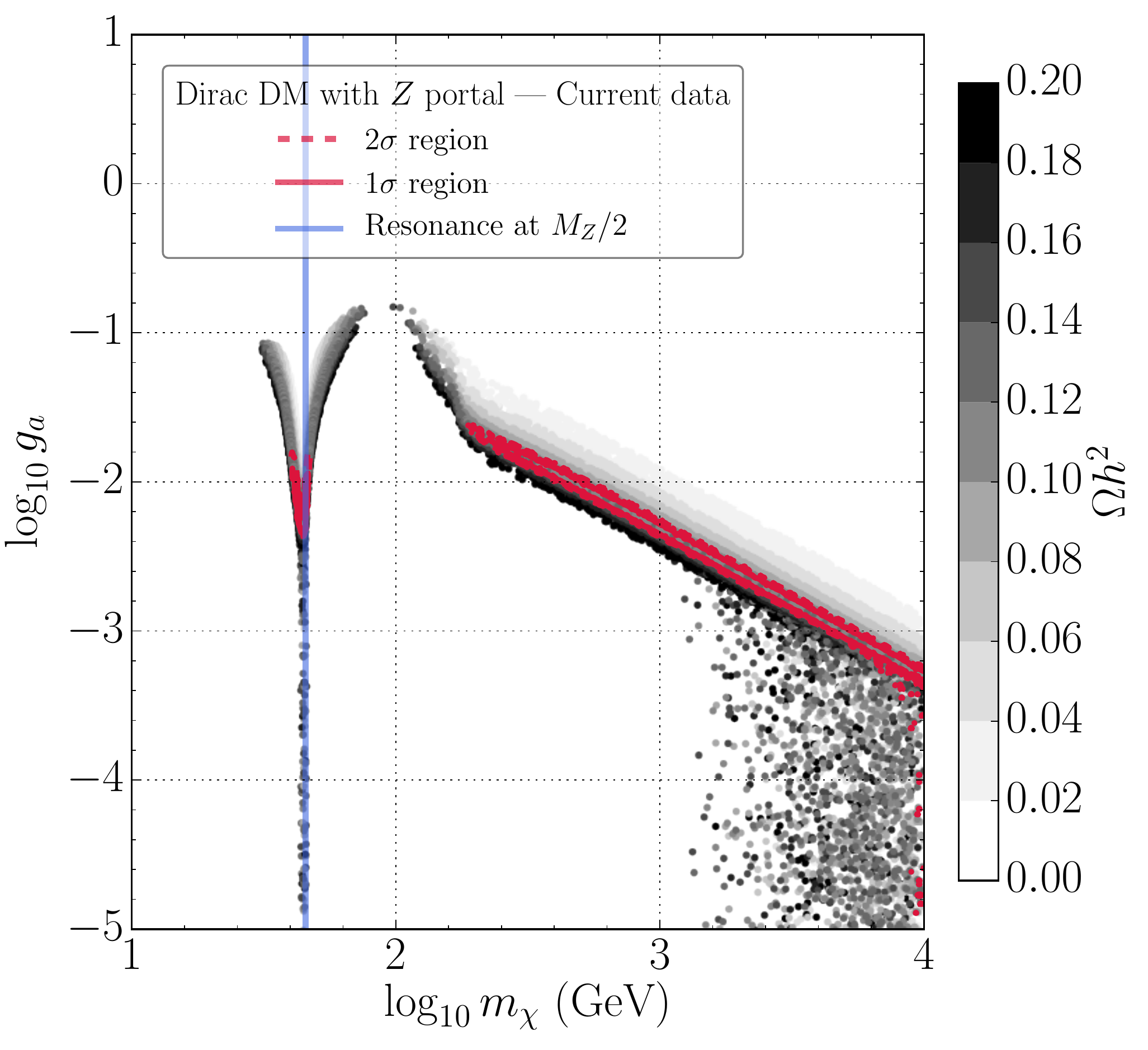}
        \caption{\it Dirac DM, axial coupling}
        \label{Fig:}
    \end{subfigure}
    \begin{subfigure}[c]{0.49\textwidth}
        \centering
        \includegraphics[width=\textwidth]{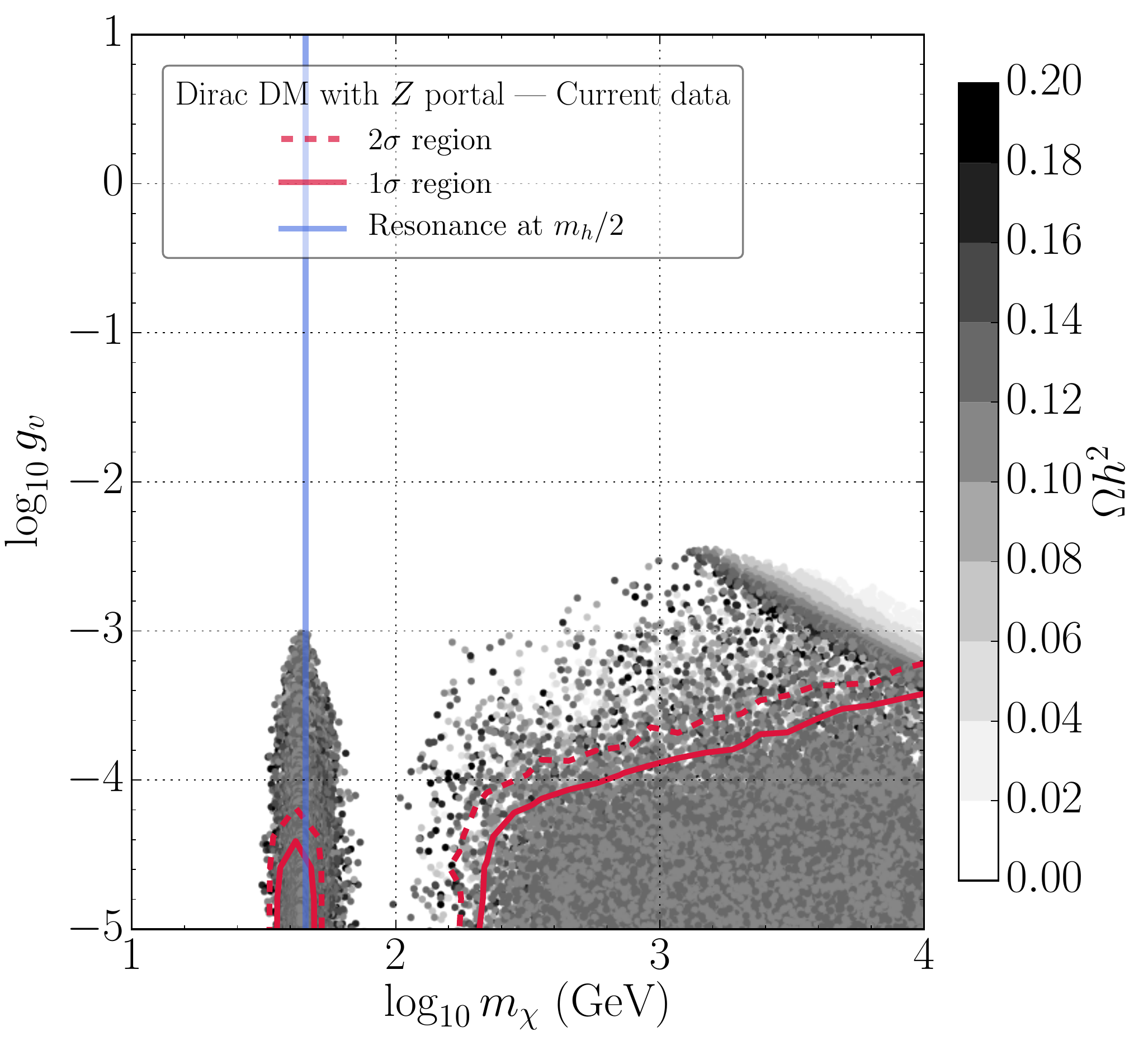}
        \caption{\it Dirac DM, vector coupling}
        \label{Fig:}
    \end{subfigure}
    \centering
    \begin{subfigure}[c]{\textwidth}
    \centering
        \includegraphics[width=0.49\textwidth]{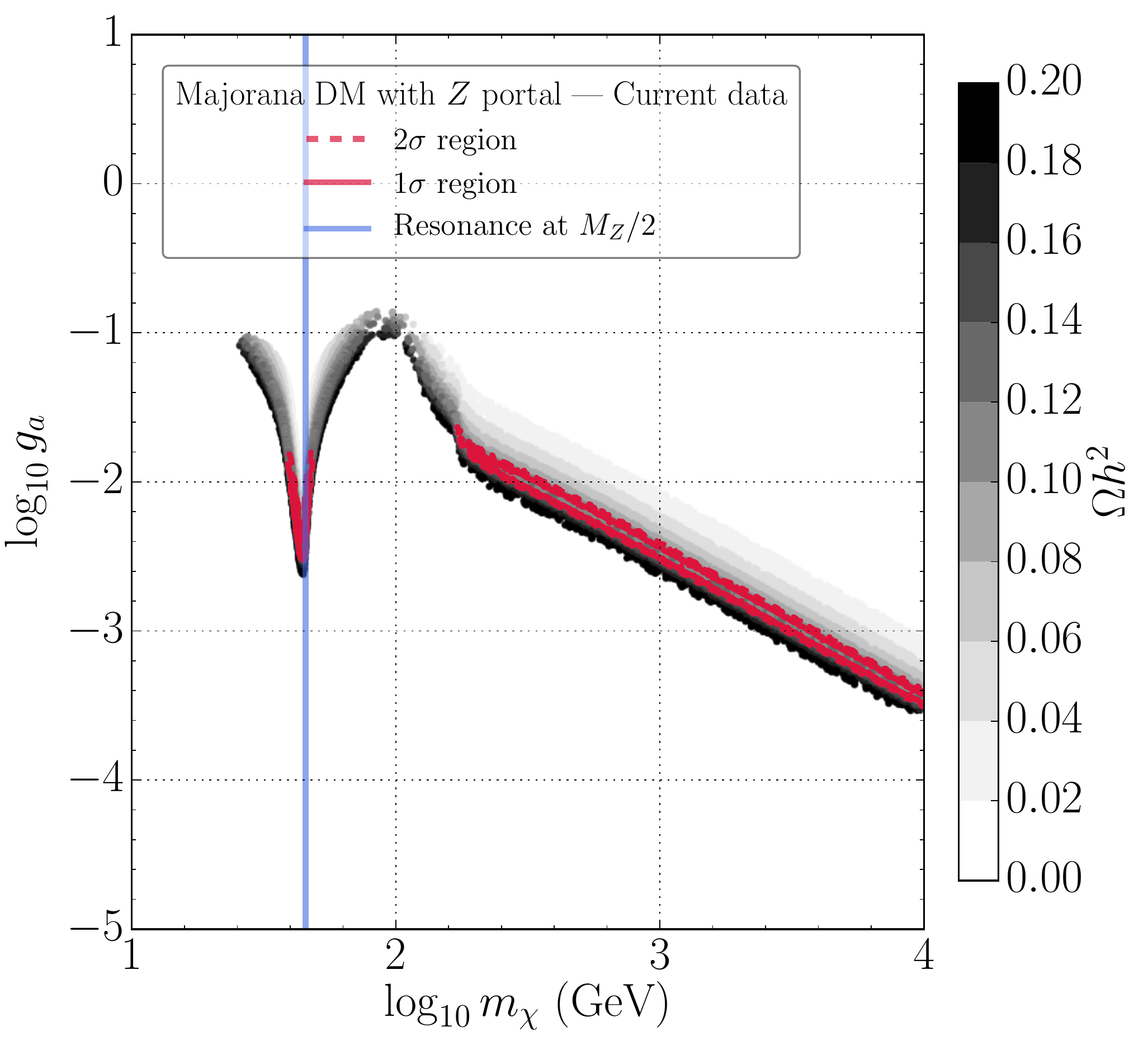}
        \caption{\it Majorana DM, axial coupling}
        \label{Fig:}
    \end{subfigure}\hfill\null
    \caption{\it Confidence regions of DM mass and coupling for $Z$-portal fermionic DM models in light of all current data.}
    \label{Fig:mass_coupling_fermionic_z_portal}
\end{figure} 

We show Bayes factors and minimum $\chi^2$ values for our SM-portal models in \reftable{Table:BF_CHI2},
considering all present data. The most successful is the Majorana $Z$-portal model, but several of the
other models have are also highly consistent with the available data, as indicated by their Bayes factors,
minimum $\chi^2$ and $p$-values. These are the $h$-portal models with scalar, vector or fermionic
DM, and the $Z$-portal models with fermionic DM. On the other hand, two of the models have calculated
Bayes factors that should be interpreted as ``decisive'' evidence against them.
These are the scalar and vector $Z$-portal models. 
Given the available data, the scalar and vector $Z$-portal models are about $10^{-14}$ 
and $10^{-10}$ less plausible than other SM-portal models.
The models are essentially falsified by DD searches, as they predict relatively large SI scattering cross sections on nucleons. 

As for the minimum $\chi^2$ values, we find that in most of the models there is sufficient freedom to minimize this quantity, barring for a small contribution from the invisible width of the $Z$-boson (the observed invisible width is slighty less than the SM prediction). This results in a common baseline for the minimum $\chi^2$ value of about $2.6$. DD however impairs the scalar and vector $Z$-portal models, inducing total $\chi^2$ values of about $54$ and $35$, respectively. With two degrees of freedom, these correspond to \pvalue{s} of about $10^{-12}$ and $10^{-8}$, respectively.

\begin{table}[ht!]
\centering
\begin{tabular}{lccc}  
\toprule
Model & Bayes factor & $\min\chi^2$ & \pvalue\\
\midrule
    Real scalar $h$-portal & \textcolor{barely}{$0.55$} & \textcolor{barely}{$2.6$} & \textcolor{barely}{$0.27$} \\ 
    Complex scalar $h$-portal & \textcolor{barely}{$0.28$} & \textcolor{barely}{$2.6$} & \textcolor{barely}{$0.27$} \\ 
    Real vector $h$-portal & \textcolor{barely}{$0.23$} & \textcolor{barely}{$2.6$} & \textcolor{barely}{$0.27$} \\ 
    Complex vector $h$-portal & \textcolor{barely}{$0.059$} & \textcolor{barely}{$2.6$} & \textcolor{barely}{$0.27$} \\ 
    Majorana $h$-portal & \textcolor{barely}{$0.59$} & \textcolor{barely}{$2.6$} & \textcolor{barely}{$0.27$} \\ 
    Dirac $h$-portal & \textcolor{barely}{$0.71$} & \textcolor{barely}{$2.6$} & \textcolor{barely}{$0.27$} \\ 
     \midrule
    Scalar $Z$-portal & \textcolor{decisive}{$3 \times 10^{-14}$} & \textcolor{decisive}{$55$} & \textcolor{decisive}{$1.4 \times 10^{-12}$} \\ 
    Vector $Z$-portal & \textcolor{decisive}{$6.8 \times 10^{-10}$} & \textcolor{decisive}{$35$} & \textcolor{decisive}{$2.2 \times 10^{-8}$} \\ 
    Majorana $Z$-portal & \textcolor{barely}{$1$} & \textcolor{barely}{$2.6$} & \textcolor{barely}{$0.27$} \\ 
    Dirac $Z$-portal & \textcolor{barely}{$0.24$} & \textcolor{barely}{$2.6$} & \textcolor{barely}{$0.27$} \\ 
\bottomrule
\end{tabular}
\caption{\it Bayes factors, $\chi^2$ values and \pvalue{}s for DM models in light of DD experiments in addition to measurements of the 
relic abundance and collider and indirect detection data. The color scheme for the Bayes factors reflects interpretation on the 
Jeffreys' scale for negative evidence~\cite{Jeffreys:1939xee}: green indicates barely worth mentioning, orange indicates substantial evidence, 
and red indicates strong, very strong or decisive evidence. The Bayes factors are relative to the best model (the Majorana $Z$-portal).
Similar colours are used for the interpretations of the \pvalue{s} calculated within the frequentist approach.}
\label{Table:BF_CHI2}
\end{table}

We display in the following figures the 
confidence regions found in the frequentist analysis of the models that were not decisively disfavoured.
The credible regions in our Bayesian analysis (not shown) are similar.
In each case, we indicate in (only 10) shades of grey the relic density calculated using \mo,
and we outline with solid (dashed) red contours the regions preferred at the 1- and 2-$\sigma$ levels.
In general, there are two regions of model parameter space of interest: one in which on-shell portal particles
can decay directly into pairs of on-shell DM particles, and another in which the DM particles can be produced
only via off-shell portal particles. As seen in Fig.~\ref{Fig:mass_coupling}, both of these are open possibilities
in the Higgs-portal model with real scalar DM. Panel (a) shows results with only the relic density and collider
constraints, panel (b) includes also the indirect DM detection constraint, panel (c) further includes the current
direct DM detection constraints, and panel (d) shows the impact of a possible null result from the LZ
experiment (the impact of a null result from the XENON1T experiment would be very similar).

The upper panels of Fig.~\ref{Fig:mass_coupling_vector_scalar_h_portal} compares the confidence regions
for Higgs-portal models with (a) real and (b) complex scalar DM, incorporating all the present constraints.
We see that the results are quite similar. We see in the lower panels of 
Fig.~\ref{Fig:mass_coupling_vector_scalar_h_portal} that the same is true for the confidence regions for Higgs-portal models 
with (c) real and (d) complex vector DM. In these models there are also two confidence regions, corresponding
to on- and off-shell Higgs couplings to the DM particles.

Fig.~\ref{Fig:mass_coupling_fermion_h_portal} displays the confidence regions for Higgs-portal models
with fermionic DM in light of all present data. The upper panels are for Dirac fermions and the lower panels for Majorana fermions.
In each case there are two couplings to the Higgs: scalar $g_s$ (left panels) and pseudoscalar $g_a$ (right panels).
There are extended regions of the scalar couplings with $\log_{10} g_s < -1.5$ that are favoured at the 1-$\sigma$ level,
whereas there are only narrow confidence bands with $\log_{10} g_p > -1.5$ in the pseudoscalar cases.
The DM may annihilate through a combination of scalar and pseudoscalar couplings. The scalar coupling is forced to be small by DD constraints, whereas the pseudoscalar coupling is not affected by DD constraints as its contributions to the scattering cross section are momentum suppressed. Since the scalar coupling must be small, the pseudoscalar coupling must set the correct relic density, and so is much more restricted. 
In these models the only confidence regions correspond to off-shell Higgs couplings to DM particles weighing $\gtrsim 100$~GeV.
The Bayes factors, minimum $\chi^2$ and $p$-values for these Higgs-portal models are
all quite comparable, as seen in Table~\ref{Table:BF_CHI2}, with insignificant evidence against any of them.

On the other hand, the same Table shows that there is decisive evidence against the scalar and vector
$Z$-portal DM models, and we do not display the mass/coupling planes for these models.
We do, however, display the corresponding planes for $Z$-portal fermionic DM models in
Fig.~\ref{Fig:mass_coupling_fermionic_z_portal}. The upper panels are for Dirac fermion DM, 
and the lower panel is for Majorana fermion DM. In the former case we consider both
axial (upper left) and axial (upper right) DM-$Z$ couplings, whereas in the Majorana case
only an axial DM-$Z$ coupling is allowed. In all three cases, only the off-shell $Z$ option with a DM particle mass
$\gtrsim 200$~GeV is credible. As seen in Table~\ref{Table:BF_CHI2}, the Majorana $Z$-portal DM model
currently has the largest Bayes factor, and also shares with the Dirac $Z$-portal case the lowest $\chi^2$ minimum 
and the highest \pvalue.

In \reftable{Table:PBF}, we show partial Bayes factors (PBFs) for DD data for SM-portal models versus a hypothetical DM model with no DD signatures. 
This illustrates the damage to DM models from DD data. We consider present DD limits from PandaX and and PICO,
projections of possible future DD limits from LZ and PICO, and the neutrino floor for spin-independent limits. The projected DD limits would
damage the plausibility of Higgs portal models. Except for the fermion portal models, the damage to their plausibility shifted from 
``barely worth mentioning'' to ``substantial.'' The status of the $Z$-portal models were unchanged, 
although the scalar and vector models were further damaged. Taking the SI limits on DD cross sections to the neutrino floor
would provide decisive evidence against all scalar and vector SM-portals if no signal were observed, but fermion portals would not be harmed.

\begin{table}[ht!]
\centering
\begin{tabular}{lccc}  
\toprule
& \multicolumn{3}{c}{Damage to plausibility from DD}\\
\cmidrule(r){2-4}
Model & Present data &  Possible Future data & Neutrino floor\\
\midrule
    Real scalar $h$-portal & \textcolor{barely}{$0.3$} & \textcolor{substantial}{$0.006$} & \textcolor{strong}{$5 \times 10^{-5}$} \\ 
    Complex scalar $h$-portal & \textcolor{barely}{$0.1$} & \textcolor{substantial}{$0.002$} & \textcolor{strong}{$1 \times 10^{-5}$} \\ 
    Real vector $h$-portal & \textcolor{barely}{$0.1$} & \textcolor{substantial}{$0.0009$} & \textcolor{vstrong}{$9 \times 10^{-7}$} \\ 
    Complex vector $h$-portal & \textcolor{substantial}{$0.02$} & \textcolor{substantial}{$0.001$} & \textcolor{decisive}{$6 \times 10^{-10}$} \\ 
    Majorana $h$-portal & \textcolor{barely}{$0.2$} & \textcolor{barely}{$0.2$} & \textcolor{barely}{$0.1$} \\ 
    Dirac $h$-portal & \textcolor{barely}{$0.2$} & \textcolor{barely}{$0.1$} & \textcolor{barely}{$0.1$} \\ 
    \midrule
    Scalar $Z$-portal & \textcolor{decisive}{$1 \times 10^{-14}$} & \textcolor{decisive}{$7 \times 10^{-73}$} & \textcolor{decisive}{$7 \times 10^{-129}$} \\ 
    Vector $Z$-portal & \textcolor{decisive}{$3 \times 10^{-10}$} & \textcolor{decisive}{$7 \times 10^{-54}$} & \textcolor{decisive}{$2 \times 10^{-101}$} \\ 
    Majorana $Z$-portal & \textcolor{barely}{$0.3$} & \textcolor{barely}{$0.2$} & \textcolor{barely}{$0.1$} \\ 
    Dirac $Z$-portal & \textcolor{barely}{$0.08$} & \textcolor{barely}{$0.04$} & \textcolor{substantial}{$0.01$} \\ 

\bottomrule
\end{tabular}
\caption{\it Partial Bayes factors (PBFs) for DM models in light of DD experiments. The PBFs show change in plausibility relative to a model unaffected by DD experiments when DD data is considered in addition to measurements of the relic abundance and collider and indirect detection data. The color scheme reflects interpretation with the Jeffreys' scale\cite{Jeffreys:1939xee}: green indicates  barely worth mentioning, orange indicates substantial evidence, and red indicates strong, very strong or decisive evidence.}
\label{Table:PBF}
\end{table}

We compare the findings from the Bayes factors with the corresponding frequentist analysis in \reftable{Table:DC}. 
The $\chi^2$ values for the scalar and vector $Z$ portals increase substantially when possible future data are considered. 
The statuses of the remaining models barely change, however, though the vector Higgs portal might be ruled out by SI limits at the neutrino floor. 
Even for our simple SM-portal models, we can fine-tune the mass and coupling to evade possible powerful constraints upon the 
DD cross section from future experiments. Fine-tuned masses and couplings can enhance DM annihilation by an $s$-channel resonance, \ie 
\begin{equation}
\sqrt{s} \approx 2 m_\chi \simeq M_Z \text{ or } m_h.
\end{equation}
Elastic scattering on nucleons, though, proceeds via $t$-channel exchange and is not enhanced. 

\begin{table}[ht!]
\centering
\begin{tabular}{lccc}  
\toprule
& \multicolumn{3}{c}{Change in $\min\chi^2$ from DD data}\\
\cmidrule(r){2-4}
Model & Present data &  Possible Future data & Neutrino floor\\
\midrule
    Real scalar $h$-portal & \textcolor{barely}{0} & \textcolor{barely}{0} & \textcolor{barely}{$0.87$} \\ 
    Complex scalar $h$-portal & \textcolor{barely}{0} & \textcolor{barely}{0} & \textcolor{barely}{$2.4$} \\ 
    Real vector $h$-portal & \textcolor{barely}{0} & \textcolor{barely}{0} & \textcolor{substantial}{$8.5$} \\ 
    Complex vector $h$-portal & \textcolor{barely}{0} & \textcolor{barely}{0} & \textcolor{vstrong}{$14$} \\ 
    Majorana $h$-portal & \textcolor{barely}{0} & \textcolor{barely}{0} & \textcolor{barely}{0} \\ 
    Dirac $h$-portal & \textcolor{barely}{0} & \textcolor{barely}{0} & \textcolor{barely}{0} \\ 
    \midrule
    Scalar $Z$-portal & \textcolor{decisive}{$52$} & \textcolor{decisive}{$3.2 \times 10^{2}$} & \textcolor{decisive}{$5.7 \times 10^{2}$} \\ 
    Vector $Z$-portal & \textcolor{decisive}{$33$} & \textcolor{decisive}{$2.3 \times 10^{2}$} & \textcolor{decisive}{$4.5 \times 10^{2}$} \\ 
    Majorana $Z$-portal & \textcolor{barely}{0} & \textcolor{barely}{0} & \textcolor{barely}{0} \\ 
    Dirac $Z$-portal & \textcolor{barely}{0} & \textcolor{barely}{0} & \textcolor{barely}{0} \\ 
\bottomrule
\end{tabular}
\caption{\it Change in $\chi^2$ for DM models in light of DD experiments in addition to measurements of the relic abundance and collider and indirect detection data. The color scheme reflects interpretation: green indicates barely worth mentioning, orange indicates substantial evidence, and red indicates strong, very strong or decisive evidence.}\label{Table:DC}
\end{table}

\begin{figure}
    \centering
    \begin{subfigure}[t]{0.49\textwidth}
        \centering
        \includegraphics[width=\textwidth]{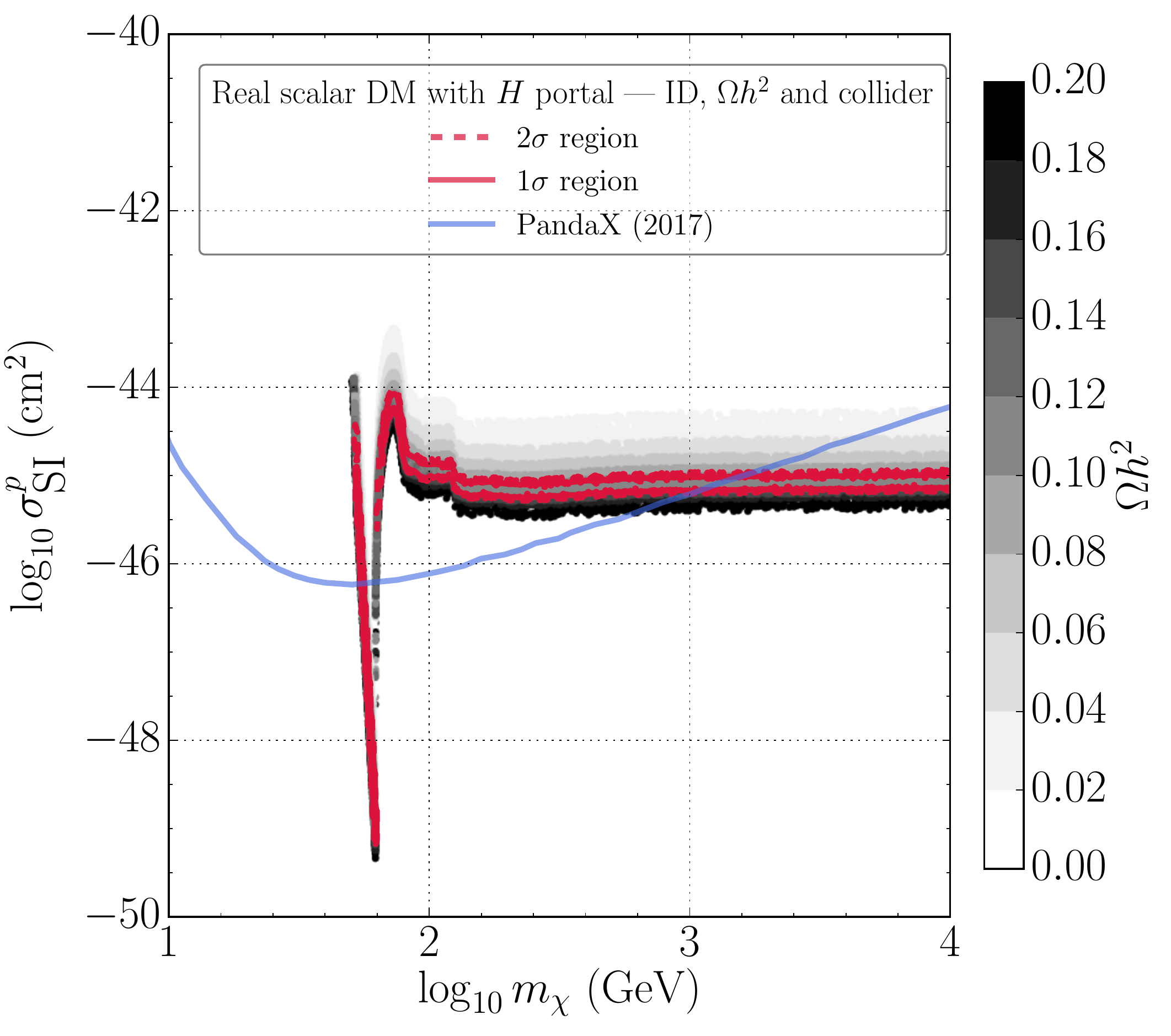}
        \caption{\it Indirect detection, relic density and \\ collider constraints.}
        \label{Fig:}
    \end{subfigure}
    \begin{subfigure}[t]{0.49\textwidth}
        \centering
        \includegraphics[width=\textwidth]{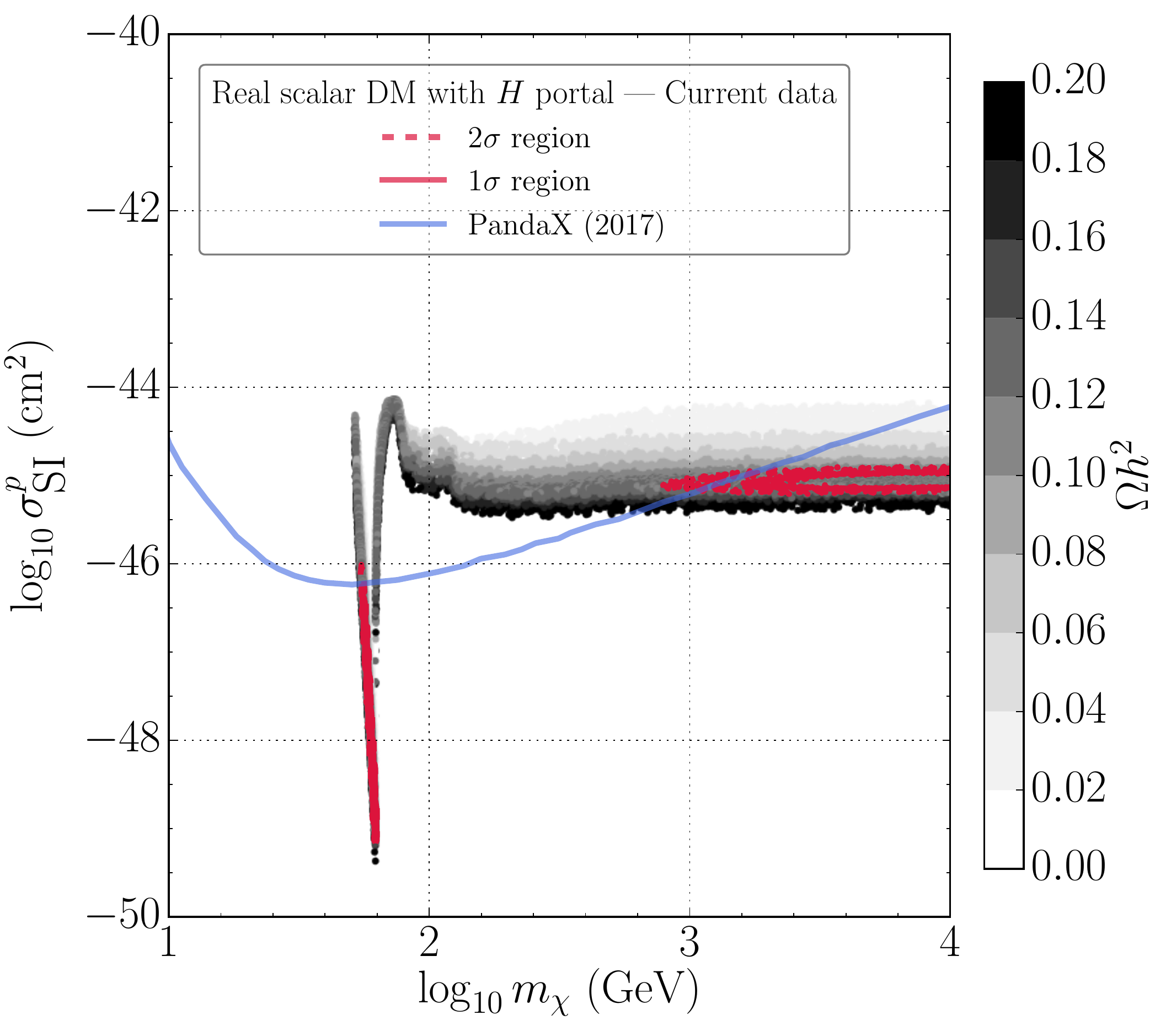}
        \caption{\it Direct and indirect detection, relic density and collider constraints.}
        \label{Fig:}
    \end{subfigure}
    \begin{subfigure}[t]{0.49\textwidth}
        \centering
        \includegraphics[width=\textwidth]{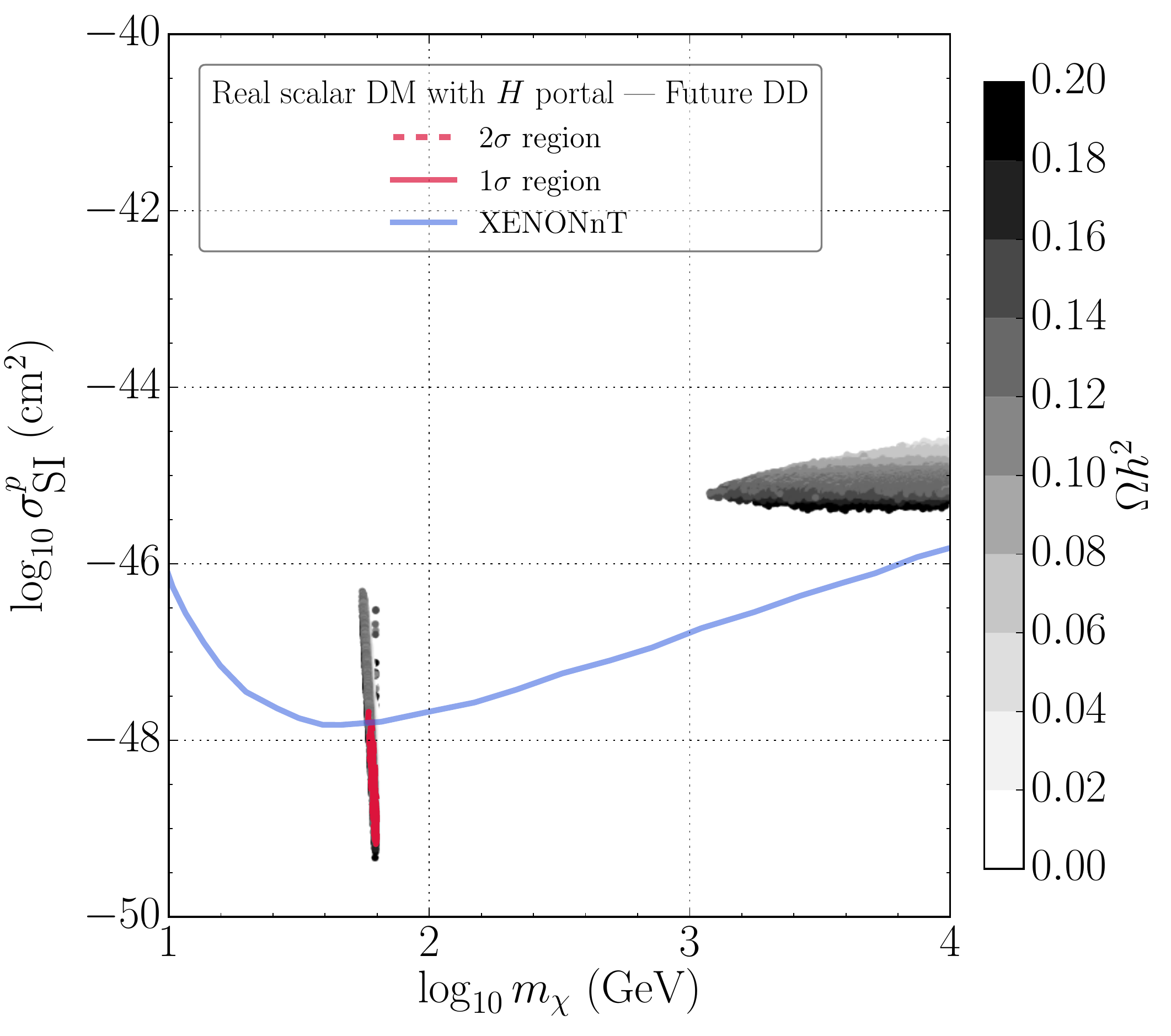}
        \caption{\it Possible LZ constraints on direct detection, \\ indirect detection, relic density and collider \\ constraints.}
        \label{Fig:}
    \end{subfigure}
    \begin{subfigure}[t]{0.49\textwidth}
        \centering
        \includegraphics[width=\textwidth]{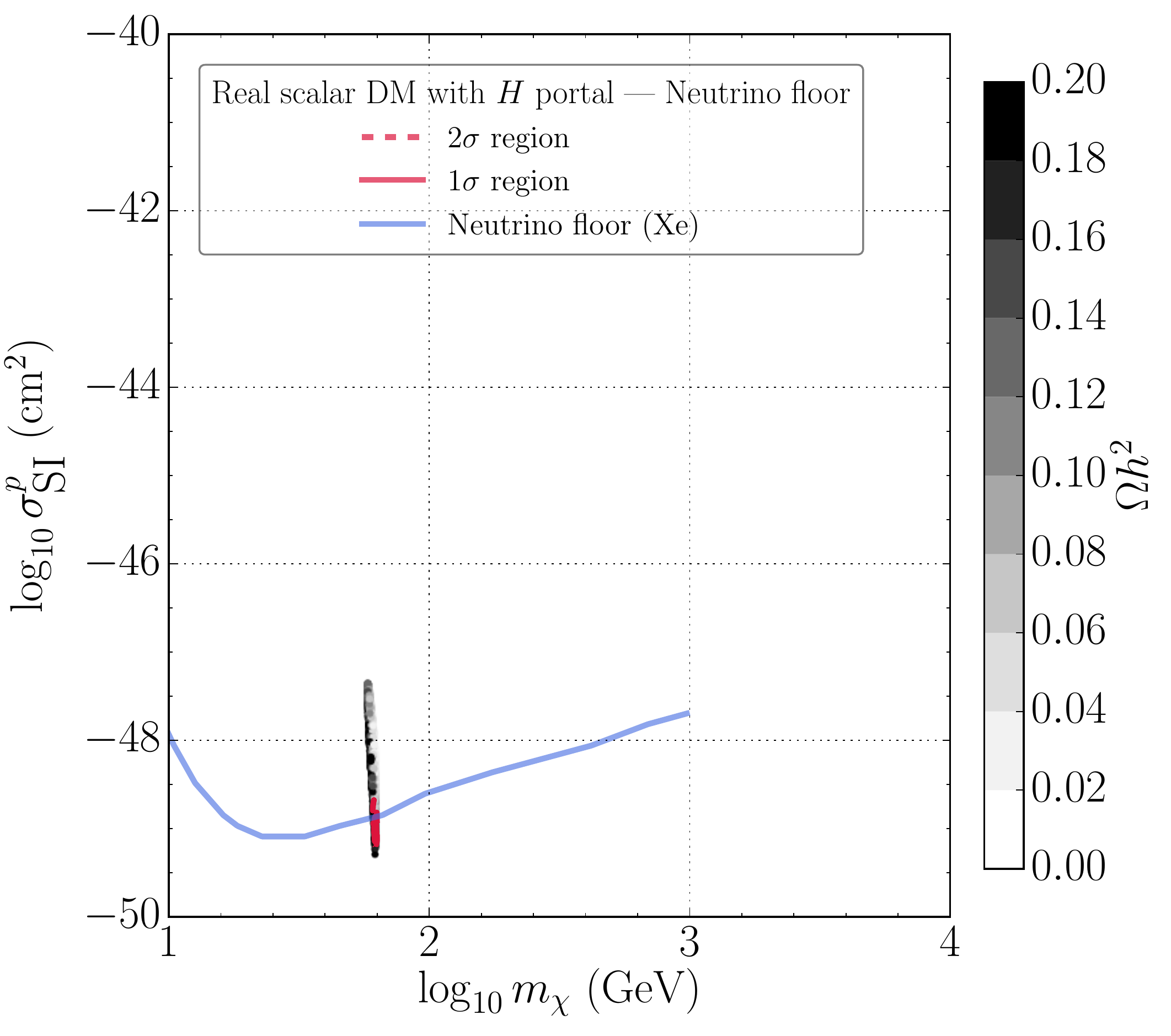}
        \caption{\it Neutrino floor limits on direct detection, indirect detection, relic density and collider constraints.}
        \label{Fig:}
    \end{subfigure}
    \caption{\it Confidence regions of DM mass and SI cross section for real scalar Higgs portal in light of data from various experiments.}
    \label{Fig:mass_sigma}
\end{figure}

\subsection{DD uncertainties}

\begin{figure}[ht!]
\centering
\includegraphics[width=0.4\textwidth]{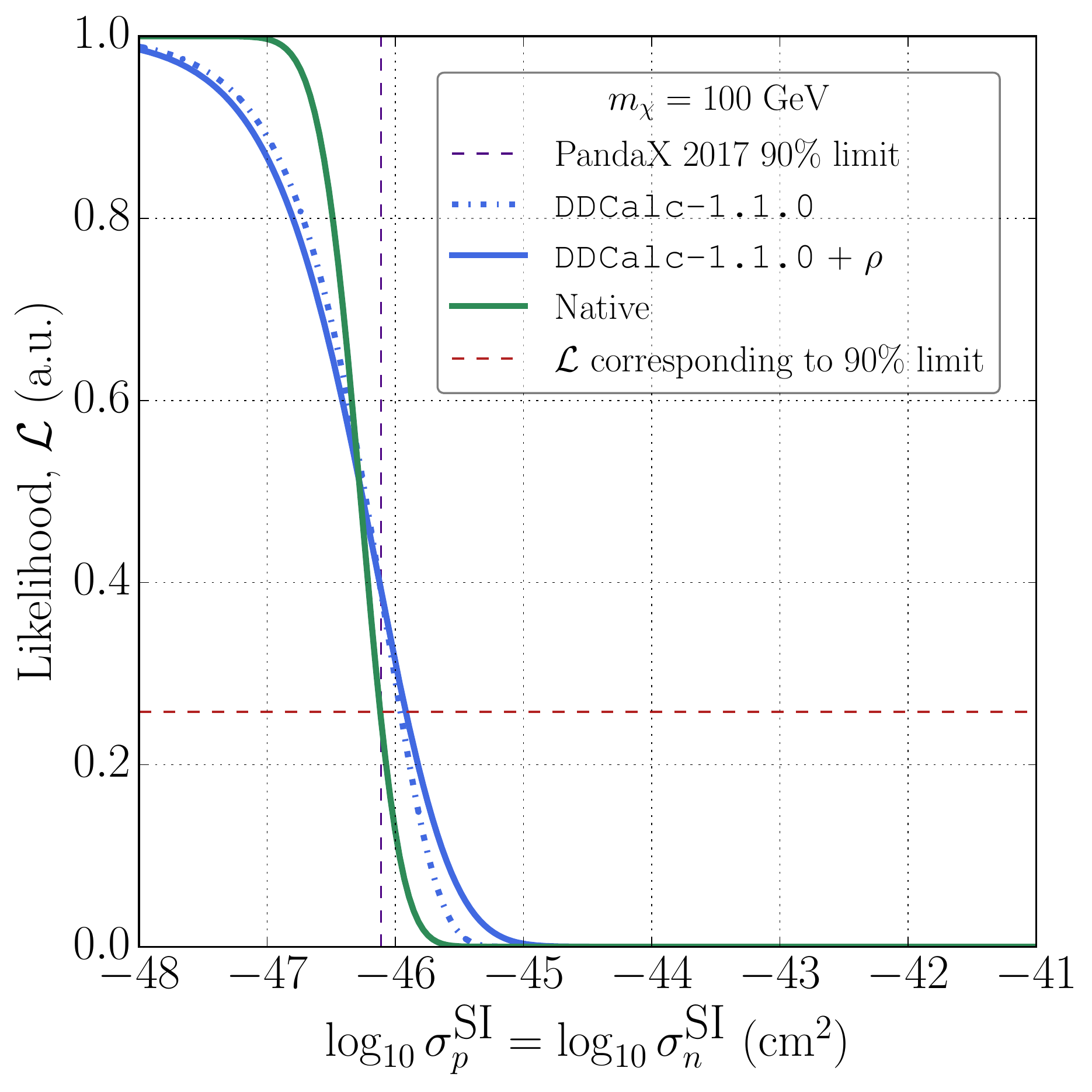}
\caption{\it Likelihood as a function of scattering cross section for PandaX with our implementation (green) and that from \ddcalc convolved with uncertainty in the local density (blue). The mass of the DM was fixed to $m_\chi = 100\gev$. Note that \ddcalc contains the fractionally weaker XENON1T 2017 data.}\label{Fig:dd_compare_1d}
\end{figure}

We find that for the impact of uncertainties in DD evidences are small. The uncertainties smear a limit in the scattering cross section, as in \reffig{Fig:dd_compare_1d}. Whilst this smearing affects the posterior density, the evidence, which is an average likelihood, is relatively stable with respect to hadronic, velocity profile, and density uncertainties. In \reftable{Table:DDCalc}, we show evidences for the real scalar Higgs-portal model with DD, ID, relic abundance and collider data, computed with our native implementation of a likelihood for DD and that from \ddcalc, with different uncertainties included. \ddcalc includes XENON1T data, which is fractionally weaker than PandaX for larger DM masses.

Uncertainties that change Bayes factors by less than about a factor of $5$ are typically considered irrelevant, since they are 
overwhelmed by changes in Bayes factors induced by changes of priors. The changes from treatments of DD uncertainties in 
\reftable{Table:DDCalc} are of order unity, and hence largely irrelevant. The biggest impact is that from including the uncertainty 
in the local density in our calculation of the likelihood, which smears a step-function to a Gaussian error function as shown in 
\reffig{Fig:dd_compare_1d}, bringing it closer towards the \ddcalc likelihood. The difference between our implementation of 
DD constraints and that in \ddcalc results in a factor of order $1$ in evidence. Including velocity profile uncertainties, which we 
omitted elsewhere, changes the evidence by a factor of about $3$, which may be considered irrelevant.

\begin{table}[ht!]
\centering
\begin{tabular}{cccccccc}  
\toprule
& \multicolumn{6}{c}{Uncertainties included in DD}\\
\cmidrule(r){2-7}
& None & $\Delta\rho$ & Had & Alt & Had + $\Delta\rho$ & Vel & Had + $\Delta\rho$ + vel\\
\midrule
    Native & $0.000589$ & $0.000996$ & $0.000484$ & $0.000462$ & $0.000396$ \\ 
    \texttt{DDCalc} & $0.000516$ & $0.000521$ & $0.000471$ & $0.000393$ & $0.000312$ & $0.000496$ & $0.000145$ \\ 
\bottomrule
\end{tabular}
\caption{\it Impact of uncertainties in evidence for real scalar Higgs-portal model with DD, ID, relic abundance and collider data. 
We compare results from our native implementation of DD likelihoods and \ddcalc; only the latter allows velocity profile uncertainties (denoted Vel). Note, however, that \ddcalc includes XENON1T data, which is fractionally weaker than PandaX for larger DM masses. 
We denote the uncertainty in the local density of DM by $\Delta\rho$ and our alternative treatment of hadronic uncertainties (Had) by Alt.}\label{Table:DDCalc}
\end{table}

\subsection{Treatment of resonances}\label{Sec:Resonance}

To insure that narrow resonances are adequately sampled, for the real scalar Higgs-portal model we investigated our sampling settings. 
We varied the number of live points, and compared multimodal and importance-nested sampling.
Furthermore, we checked that we obtained similar results with an alternative technique in which we traded a coupling 
for the relic density, $g\to\Omega$, such that
\begin{equation}
\ev = \int \mathcal{L}(\Omega) \mathcal{L}(\cdots) \left.\pi(g_\Omega, \cdots) \frac{\pd g}{\pd \Omega}\right|_{g_\Omega} \,d\Omega d\cdots \, .
\end{equation}
We performed the resulting integral by sampling the relic density from a Gaussian likelihood, $\mathcal{L}(\Omega)$, which played the role of a prior, 
solving for the coupling as a function of the sampled relic density, $g_\Omega$, by Brent's method and calculating the derivative, 
which played the role of a likelihood. We found no significant changes in our evidences.
To calculate confidence regions, we furthermore performed dedicated scans around the resonance to insure that it was adequately sampled.

\subsection{Prior sensitivity}\label{Sec:sensitivity}

In our Bayesian analysis, we find that with linear priors for the DM mass and coupling the evidence is about 10 times greater than that with 
logarithmic priors for the real scalar Higgs portal with current data and with current data except direct detection experiments. 
This can be ascribed to the fact that linear priors favor larger values for the DM mass, 
where experimental constraints from direct searches are weaker. A factor of ten would not alter qualitative conclusions 
drawn from our Bayes factors in \reftable{Table:PBF}, as Bayes factors are typically interpreted on a logarithmic scale. 

For the Majorana $Z$-boson portal, we find evidences about $100$ times smaller with linear priors than with 
logarithmic priors with current data and also when the direct detection experiments are omitted. This arises because
a small coupling, as required by this model, is disfavoured by a linear prior. Factors of 100 may alter qualitative conclusions 
drawn from the Bayes factors in \reftable{Table:BF_CHI2}. This is not surprising: with qualitatively different prior beliefs 
about the scales of couplings and mass of DM, one may reach qualitatively different conclusions about the relative plausibility of 
SM-portal models. However, our default is to present evidences from logarithmic priors,
since we are agnostic about the order of magnitude of the DM mass and couplings whereas linear priors, 
on the other hand, favour the greatest permitted orders of magnitude.

The partial Bayes factors in \reftable{Table:PBF} (which involve ratios of the aforementioned evidences) 
are rather insensitive to the choice of logarithmic or linear priors, suggesting that conclusions drawn from them are somewhat robust. 
The frequentist $\chi^2$ results, which we compare with the Bayesian analysis, are of course insensitive to our choice of prior.

\subsection{Number of constrained parameters}

Following \refcite{RSSB:RSSB353,Kunz:2006mc,Martin:2013nzq}, we consider the effective number of constrained parameters in each model
\begin{equation}
n_\text{eff} = \langle \chi^2\rangle - \min\chi^2,
\end{equation}
where $\langle \cdot \rangle$ denotes a posterior mean. This heuristic for the number of constrained parameters originates from considering the 
Kullback-Leibler divergence between the prior and posterior probability density functions.

To see that this is a heuristic for the number of constrained parameters, imagine an $n$-dimensional linear model in which the 
likelihood is a product of Gaussians with widths $\sigma_i$ for each of $n$ parameters and the priors are uniform with widths 
$w_i$. If the parameters are constrained, \ie $\sigma_i \ll w_i$, the posterior is approximately Gaussian. 
Thus $\langle \chi^2\rangle$ is the mean of a $\chi^2$ distribution with $n$ degrees of freedom, \ie $\langle \chi^2\rangle = n$. 
The minimum $\chi^2 = 0$; thus $n_\text{eff} = n$. 

If, on the other hand, $m$ parameters are unconstrained, \ie $\sigma_i \gg w_i$, their posterior is approximately uniform. 
Thus their contributions to $\langle \chi^2\rangle$ are averaged upon a uniform distribution and approximately vanish such that 
$\langle \chi^2\rangle \approx n - m$ and we find $n_\text{eff} \approx n - m$. This example demonstrates that whether a parameter 
should be considered constrained depends upon our prior.

The effective number of parameters for each model is shown in \reftable{Table:Neff}. We find that the effective numbers of 
parameters range from about $1$ -- $3$, as expected. \refcite{Kunz:2006mc} argues that when evidences are similar, 
we should discriminate between models by their numbers of constrained parameters, favouring models with fewer. 
While not necessarily endorsing this point of view, we note that we find that Higgs-portal models have fewer constrained parameters than $Z$-portal models.

\begin{table}[ht!]
\centering
\begin{tabular}{ll}  
\toprule
Model & $n_\text{eff}$\\
\midrule
    Real scalar $h$-portal & $1.6$ \\ 
    Complex scalar $h$-portal & $2$ \\ 
    Real vector $h$-portal & $2.6$ \\ 
    Complex vector $h$-portal & $4.1$ \\ 
    Majorana $h$-portal & $0.99$ \\ 
    Dirac $h$-portal & $0.98$ \\ 
    \midrule
    Scalar $Z$-portal & $3.3$ \\ 
    Vector $Z$-portal & $3.3$ \\ 
    Majorana $Z$-portal & $1.4$ \\ 
    Dirac $Z$-portal & $1.5$ \\ 

\bottomrule
\end{tabular}
\caption{\it Effective number of constrained parameters in each model for DD, ID, relic abundance and collider data,
as calculated using the prescription of~\cite{RSSB:RSSB353,Kunz:2006mc,Martin:2013nzq}.}\label{Table:Neff}
\end{table}

\section{Conclusions}\label{Sec:conclusions}

We have presented Bayesian and frequentist appraisals of models with DM of spin 0, 1/2 or 1 that interacts
with SM particles and annihilates via interactions with the SM Higgs or $Z$-boson, in light of constraints from Planck, DD, ID and LHC experiments. 
We have also considered the possible impacts of null results from future DD searches at LZ and PICO.
The Bayesian and frequentist analyses yield similar conclusions, and our results are relatively insensitive to the
uncertainties in the DD scattering matrix elements, and to the choice of Bayesian priors.

We find that all the Higgs-portal models studied are compatible with the available data, and that they offer prospects
for on-shell decays of the Higgs boson into pairs of DM particles, as well as allowing for the possibility of heavier
DM particles that can be produced only via off-shell Higgs bosons. In the case of $Z$-portal models, we find that 
the available data already provide decisive evidence against the spin-0 and spin-1 cases, though they do allow both
the Majorana and Dirac spin-1/2 fermion options. However, in these cases only off-shell $Z$ interactions are
allowed. Null results from future DD experiments would provide substantial evidence against the scalar and
vector Higgs-portal models, but the fermionic Higgs- and $Z$-portal models would still be viable. Null results
of DD experiments down to neutrino `floor' levels would provide decisive evidence against the scalar and
vector Higgs-portal models, and start to provide substantial evidence against the Dirac $Z$-portal model.

We argue that our statistical analyses and our investigation of uncertainties in DD may be the most comprehensive to date. 
It is for this reason that our results differ in emphasis from some of the previous literature. The underlying Lagrangian models we study are similar to those used in previous papers, we use a similar set of phenomenological constraints, and our implementations are also similar. However, we consider the entire parameter spaces, rather than slices, and our combination of statistical approaches enables quantitative and precise characterizations of the allowed parameter spaces for all the models we study, permitting more complete statements about the regions of parameter space in which they may survive, as well as how they could be probed in the future.

\section*{Acknowledgements}
The work of J.E. was supported in part by STFC (UK) via the research grant ST/L000326/1 and in part by the Estonian Research Council via a Mobilitas Plus grant MOBTT5. The work  of L.M. and M.R. was supported by Estonian Research Council grant IUT23-6 and by the ERDF Centre of Excellence project TK133.

\appendix

\section{Statistics}\label{Sec:stats}

We analyze SM-portal models with Bayesian and frequentist statistics. Within each framework, we require two distinct methods: a method to estimate the parameters in a DM model, \eg the DM mass; and a method to judge the viability of the DM model. An ingredient in all calculations is the likelihood --- the probability of obtaining experimental data $D$ given a parameter point $x$, in a model $M$, \ie
\begin{equation}
\mathcal{L}(x) \equiv \pg{D}{x,  M}.    
\end{equation}
The likelihood in our analysis is a product of statistically independent likelihoods from measurements of the DM abundance and DD, ID and LHC experiments, 
\begin{equation}
\like(x) = \like_\text{Planck}(x) \times \like_\text{ID}(x) \times \like_\text{DD}(x) \times \like_\text{LHC}(x).
\end{equation}
The individual likelihoods are described in \refsec{Sec:like}.

For parameter inference with frequentist statistics, we calculate confidence regions by Wilks' theorem. For a two-dimensional confidence region for parameters $a$ and $b$ in a model with parameters $x = \{a, b, c\}$, by Wilks' theorem,
\begin{equation}
\Delta \chi^2(a, b) \equiv -2 \ln \frac{\like(a, b, \hat c)}{\like(\hat x)} \sim \chi^2_2,
\end{equation}
where $\hat x$ are the best-fitting parameters and $\hat c$ is the best-fit $c$ for a particular $a$ and $b$. The, \eg $95\%$ confidence interval is the region in $(a, b)$ in which $\Delta \chi^2(a, b) \le 5.99$. For judging model viability with frequentist statistics, we construct a test statistic
\begin{equation}
\text{minimum $\chi^2$} \equiv -2 \ln \frac{\like(\hat x)}{\like_\text{max}}.
\end{equation}
Unfortunately, as the distribution of the above quantity is unknown, it is not possible to calculate precisely a \pvalue, \ie the probability of obtaining a test-statistic so extreme were the null hypothesis true. See, \eg the discussion of the difficulties in estimating the distribution in a similar analysis in \refcite{Athron:2017kgt}. For illustration, though, we calculate \pvalue{}s assuming a $\chi^2$ distribution with two degrees of freedom, $\text{minimum $\chi^2$} \sim \chi^2_2$, which may be reasonable given the constraints and parameters in our models, and the predictions of our DM models at their best-fit points.

Our Bayesian methodology requires a further ingredient: priors, $\pg{x}{M}$, for the parameters $x$ in a model $M$. The priors for our SM-portal models are discussed in \refsec{Sec:models}. We update the priors with experimental data by Bayes' theorem, resulting in posterior distributions
\begin{equation}
\pg{x}{D, M} = \frac{\like(x) \cdot \pg{x}{M}}{\pg{D}{M}}.
\end{equation}
For parameter inference, we marginalize parameters that are not of interest, \eg
\begin{equation}
\pg{a, b}{D, M} = \int \pg{a, b, c}{D, M} dc \, ,
\end{equation}
and find so-called credible regions: regions of $(a, b)$ that contain a particular fraction \eg $95\%$ of the posterior. The regions are not unique; we use the ordering rules detailed in \refcite{Fowlie:2016hew}. For judging the viability of a model, we calculate directly its change in relative plausibility\cite{Jeffreys:1939xee}. This involves calculating evidence integrals,
\begin{equation}
\pg{D}{M} = \int dx \like(x) \pg{x}{M}. 
\end{equation}
The change in relative plausibility of models $M_a$ and $M_b$ in light of data $D$ may be found by Bayes' theorem,
\begin{equation}
\frac{\dfrac{\pg{M_a}{D}}{\pg{M_b}{D}}} 
{\dfrac{p(M_a)}{p(M_b)}}
= \frac{\pg{D}{M_a}}{\pg{D}{M_b}} 
\end{equation}
This may be expressed in words as
\begin{equation}
\text{Change in relative plausibility} = \frac{\text{Posterior odds}}{\text{Prior odds}}  = \text{Bayes factor}
\end{equation}
Thus we calculate Bayes factors, which are ratios of evidences, to judge changes in the relative plausibility of SM-portal models in light of data.

A merit of the Bayes factor, which is particularly relevant to our analysis of DM models, is that we may judge damage to the plausibility of a model caused by experiments that disfavour a model without excluding its entire parameter space. With \pvalue{}s, the status of a DM model would change only once the entirety of its parameter space was excluded by \eg DD experiments. With Bayes factors, however, the plausibility of a DM model would deteriorate in light of failed searches, even if there remained parameter points that are not excluded.

\begin{figure}
    \centering
    \begin{subfigure}[c]{0.49\textwidth}
        \centering
        \includegraphics[width=\textwidth]{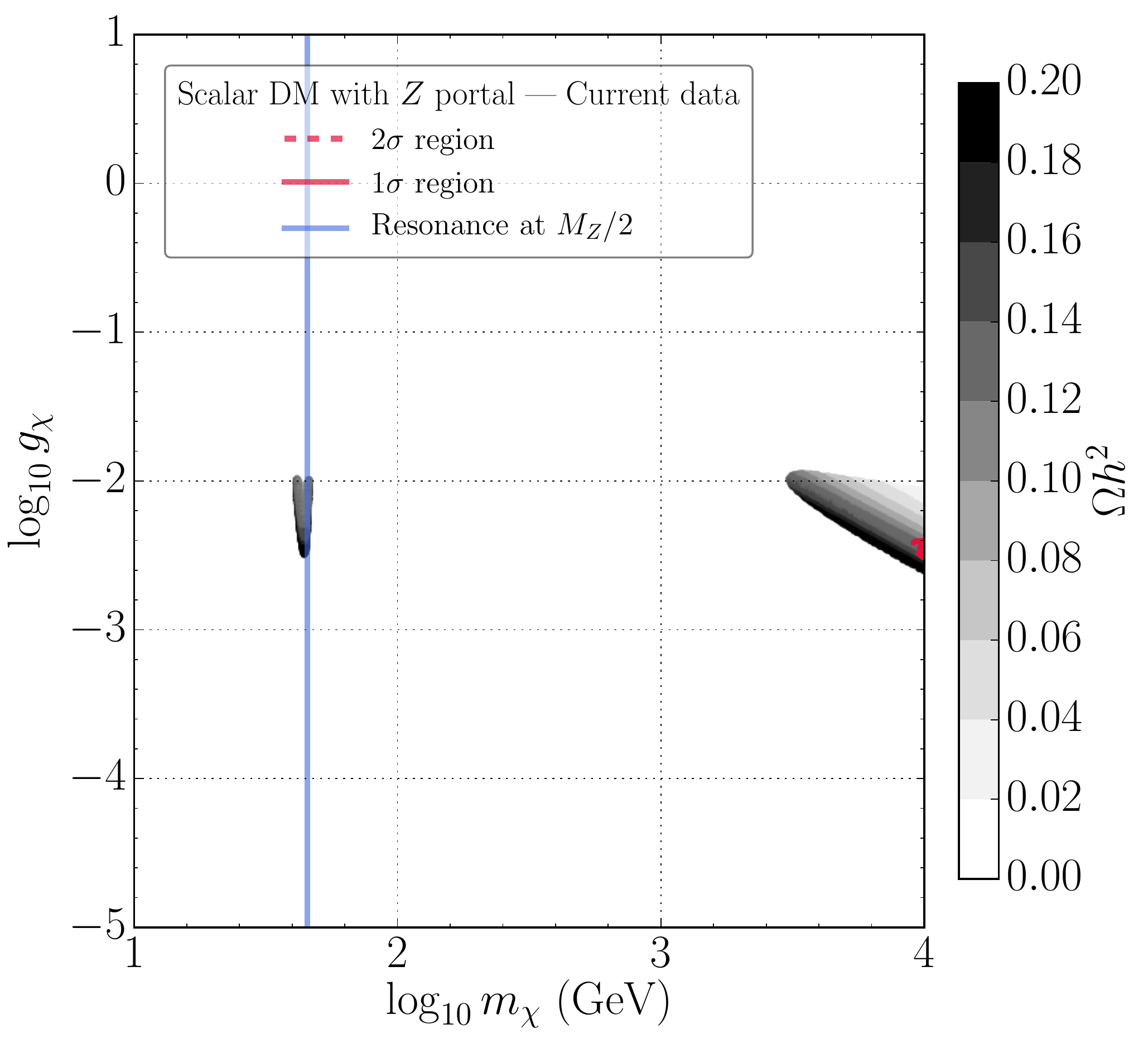}
        \caption{\it }
        \label{Fig:}
    \end{subfigure}
    \begin{subfigure}[c]{0.49\textwidth}
        \centering
        \includegraphics[width=\textwidth]{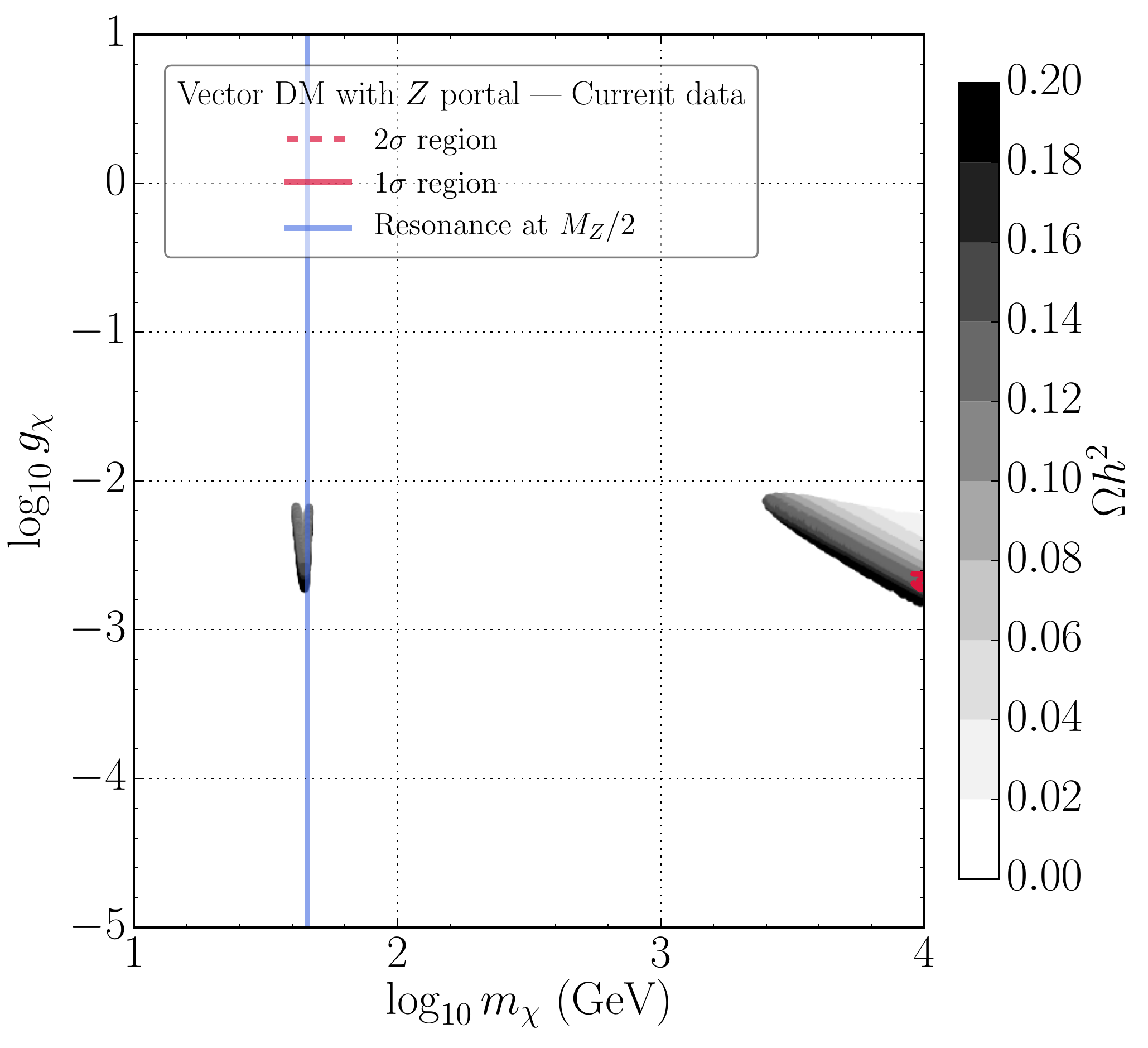}
        \caption{\it }
        \label{Fig:}
    \end{subfigure}
    \caption{\it Confidence regions of DM mass and coupling for $Z$ portal scalar and vector DM models in light of all current data.}
    \label{Fig:mass_coupling_vector_scalar_z_portal}
\end{figure}

\begin{table}[ht]
\centering
\begin{tabular}{llll}  
\toprule
\multicolumn{4}{l}{Relic density}\\
\midrule
$\Omega h^2$ & $0.1199 \pm 0.0022 \pm 10\%$ & Gaussian & Planck\cite{Ade:2015xua}\\
\midrule
\multicolumn{4}{l}{Invisible widths}\\
\midrule
$\Gamma_Z^\text{inv}$ & $499.0\pm1.5\pm0.014\mev$ & Gaussian & LEP, PDG combination\cite{Olive:2016xmw}\\
$\BR_h^\text{inv}$ & $\lesssim0.24$ & Likelhood was published & CMS\cite{CMS-PAS-HIG-16-016}\\
\midrule
\multicolumn{4}{l}{Direct detection}\\
\midrule
$\sigma_\text{SI}^{p, n}$ & $\lesssim 10^{-46}\text{cm}^2$ & $90\%$ limit $\otimes\, \Delta \rho_\text{DM}$ & PandaX\cite{Cui:2017nnn}\\
$\sigma_\text{SD}^p$ & $\lesssim 10^{-40}\text{cm}^2$ & $90\%$ limit $\otimes\, \Delta \rho_\text{DM}$ &  PICO-60\cite{Amole:2017dex}\\
$\sigma_\text{SD}^n$ & $\lesssim 10^{-40}\text{cm}^2$ & $90\%$ limit $\otimes\, \Delta \rho_\text{DM}$ & PandaX\cite{Fu:2016ega} \\ 
\midrule
\multicolumn{4}{l}{Projected direct detection}\\
\midrule
$\sigma_\text{SI}^{p, n}$ & $\lesssim 10^{-47}\text{cm}^2$ & $90\%$ limit $\otimes\, \Delta \rho_\text{DM}$ & XENONnT projection\cite{Aprile:2015uzo}\\
$\sigma_\text{SD}^p$ & $\lesssim 10^{-42}\text{cm}^2$ & $90\%$ limit $\otimes\, \Delta \rho_\text{DM}$ & PICO-500 projection\cite{pico}\\
$\sigma_\text{SD}^n$ & $\lesssim 10^{-42}\text{cm}^2$ & $90\%$ limit $\otimes\, \Delta \rho_\text{DM}$ & LZ projection\cite{Cushman:2013zza}\\
\midrule
\multicolumn{4}{l}{Spin-independent direct detection floor from coherent neutrino scattering backgrounds}\\
\midrule
$\sigma_\text{SI}^{p, n}$ & $\lesssim 10^{-49}\text{cm}^2$ & Discovery limit $\otimes\, \Delta \rho_\text{DM}$ & Xe, Ruppin et al.\cite{Ruppin:2014bra}\\
$\sigma_\text{SD}^{p}$ & $\lesssim 10^{-43}\text{cm}^2$ & Discovery limit $\otimes\, \Delta \rho_\text{DM}$ & C$_3$F$_8$, Ruppin et al.\cite{Ruppin:2014bra}\\
$\sigma_\text{SD}^{n}$ & $\lesssim 10^{-43}\text{cm}^2$ & Discovery limit $\otimes\, \Delta \rho_\text{DM}$ & Xe, Ruppin et al.\cite{Ruppin:2014bra}\\
\midrule
\multicolumn{4}{l}{Indirect detection --- $u\bar u$, $b\bar b$, $WW$, $e\bar e$, $\mu\bar\mu$ and $\tau\bar\tau$ channels}\\
\midrule
$\langle \sigma v \rangle$ & $\lesssim 10^{-26}\text{cm}^3/\text{s}$ &  $95\%$ limit $\otimes$ $J$-factors & Fermi-LAT dSphs six-year\cite{Ackermann:2015zua}\\
\midrule
\multicolumn{4}{l}{LHC --- see \reftable{Table:LHC}}\\
\bottomrule
\end{tabular}
\caption{\it Experimental data and projected data in our scans. The symbol $\otimes$ denotes a convolution of a limit with an uncertainty, \eg  $\otimes\, \Delta \rho_\text{DM}$ denotes
convolution with uncertainty in the local density of DM.}\label{Table:Data}
\end{table}
\begin{table}[ht!]
\centering
\begin{tabular}{llll}  
\toprule
Parameter & Range & Prior &\\
\midrule
$m_\chi$& $1\gev$ -- $10\tev$ & Log\\
$g$ & $10^{-6}$ -- $4\pi$ & Log\\
\midrule
Nuclear nuisance & Data & & Prior\\
\midrule
$\sigma_s$ & $41.1\pm8.1^{+7.8}_{-5.8}\mev$ & Lattice, ETM\cite{Abdel-Rehim:2016won} & Gaussian\\
$\sigma_{\pi N}$ & \hspace{-0.3cm}\rdelim\{{2}{9pt} $37.2\pm2.6^{+4.7}_{-2.9}\mev$ & Lattice, ETM\cite{Abdel-Rehim:2016won}\rdelim\}{2}{9pt} & Flat + tails; see text\\
& $58\pm5\mev$ & Pheno\cite{RuizdeElvira:2017stg}\\ 
$m_u / m_d$ & $0.38$ -- $0.58$ & Lattice, PDG comb.\cite{Olive:2016xmw} & Flat\\
$m_s / m_d$ & $17$ -- $22$ & Lattice, PDG comb.\cite{Olive:2016xmw} & Flat\\
$\Delta u + \Delta \bar u$ & $0.842\pm0.004\pm0.008\pm0.009$ & HERMES\cite{Airapetian:2006vy} & Fixed \\
$\Delta d + \Delta \bar d$ & $-0.427\pm0.004\pm0.008\pm0.009$ & HERMES\cite{Airapetian:2006vy} & Fixed \\
$\Delta s + \Delta \bar s$ & $-0.085\pm0.013\pm0.008\pm0.009$ & HERMES\cite{Airapetian:2006vy} & Fixed \\
$\delta u$ & $0.839 \pm 0.060$ & Lattice\cite{Aoki:1996pi} & Fixed\\
$\delta d$ & $-0.231\pm 0.055$ & Lattice\cite{Aoki:1996pi} & Fixed\\
$\delta s$ & $-0.046 \pm 0.034$ & Lattice\cite{Aoki:1996pi} & Fixed\\
\midrule
\multicolumn{4}{l}{Astrophysical nuisance}\\
\midrule
$\rho_\text{DM}$  & $0.3\gev/\text{cm}^3$ & Log-normal, $\sigma_{\ln \rho} = \ln 2$ \\
$J$-factor for dSphs & Log-normal, $\sigma_{\log_{10} J} = 0.25$\cite{Ackermann:2015zua}\\
\midrule
\multicolumn{4}{l}{Standard halo --- see text; included in single scan as cross-check. Distributions as in \refcite{Workgroup:2017lvb}}\\
\midrule
$v_\text{esc}$ & $550\pm35\,\text{km/s}$ & Gaussian\\
$v_\text{rel}$ & $235 \pm 20\,\text{km/s}$ & Gaussian \\
$v_0$ &  $235\pm20\,\text{km/s}$ & Gaussian\\
\midrule
\multicolumn{4}{l}{SM nuisance}\\
\midrule
$M_Z$ & $91.1876\pm0.0021\gev$ & Gaussian & LHC, PDG combination\cite{Olive:2016xmw}\\
$m_h$ & $125.09\pm0.24\gev$ & Gaussian & LEP, PDG combination\cite{Olive:2016xmw}\\
\bottomrule
\end{tabular}
\caption{\it Priors for model and nuisance parameters in our scans. For details of the nuclear matrix elements, see \refsec{Sec:DD}.}\label{Table:Priors}
\end{table}

\FloatBarrier

\bibliography{DM.bib}
\bibliographystyle{hunsrt}

\end{document}